\documentclass[aps,prc,reprint,amsmath,amssymb,amsfonts,floatfix]{revtex4-2}
\pdfoutput=1
\usepackage{latexsym,array,bm}
\usepackage{graphicx}
\usepackage{multirow,dcolumn}
\usepackage[pdftex,
            pdfauthor={Carl R. Brune},
            pdftitle={},
            pdfsubject={},
            pdfkeywords={},
            pdfproducer={revtex4-2, pdflatex with hyperref},
            pdfcreator={revtex4-2, pdflatex with hyperref}]{hyperref}

\begin{document}

\title{Spectroscopic factors, overlaps, and isospin symmetry from an
  {\em R}-matrix point of view}
\author{Carl~R.~Brune}
\email{brune@ohio.edu}
\affiliation{%
  Edwards Accelerator Laboratory \\
  Department of Physics and Astronomy \\
  Ohio University, Athens, Ohio 45701, USA }

\date{\today}

\begin{abstract}
\begin{description}
\item[Background]
Spectroscopic factors, overlaps, and isospin symmetry are often used
in conjunction with single-particle wave functions for the
phenomenological analysis of nuclear structure and reactions.
Many differing prescriptions for connecting these quantities to
physically relevant asymptotic normalization constants or widths
are available in the literature, but their relationship and
degree of validity are not always clear.
\item[Purpose]
This paper derives relationships among the above quantities of interest
using well-defined methodology and starting assumptions.
\item[Method]
$R$-matrix theory is used as the primary tool to interoperate between
the quantities of interest to this work.
Particular attention is paid to effects arising from beyond the nuclear
surface, where isospin symmetry is strongly violated.
\item[Results]
Relationships among the quantities of interest are derived.
Example applications of these methods to mirror levels in
nucleon~+~${}^{12}{\rm C}$, nucleon~+~${}^{16}{\rm O}$, and
nucleon~+~${}^{26}{\rm Al}$ are presented.
A new approach to multi-level mirror symmetry is derived and applied to
the first three $2^+$ states of ${}^{18}{\rm O}$ and ${}^{18}{\rm Ne}$.
\item[Conclusions]
The relationship between the quantities of interest is clarified and
certain procedures are recommended.
It is found that the asymptotic normalization constant of the second
$2^+$ state in ${}^{18}{\rm Ne}$ deduced from the mirror state in
${}^{18}{\rm O}$ is significantly larger than found in previous work.
This finding has the effect of increasing the
${}^{17}{\rm F}(p,\gamma){}^{18}{\rm Ne}$ reaction rate in novae.
\end{description}
\end{abstract}

\maketitle

\section{Introduction}

The concepts of spectroscopic factors, overlaps, and isospin
symmetry are widely used for the phenomenological analysis and
conceptual understanding of nuclear structure and reactions.
These quantities are often used as a ``black box,''
with little understanding of how they relate to each other or to
more fundamental descriptions of nuclei.
These quantities have physical counterparts, asymptotic normalization
constants (ANCs) and widths, that are the relevant ones in
experiments and applications.
$R$-matrix theory provides a convenient framework for unifying these
descriptions.

This work also focuses on energies near nucleon separation thresholds,
where significant effects due to the continuum may arise.
This energy regime is also a pertinent one for understanding thermonuclear
reaction rates in astrophysics, where the use of these concepts
has recently been discussed~\cite{Bar16,Bru15}.
Most of these methods have been developed for many decades, but the
results are scattered throughout the literature and in some instances forgotten.
There is presently a resurgence of interest in these principles, due to
the interest in astrophysical applications and the availability of
new radioactive ion beams. In many cases, the determination of astrophysical
reaction rates requires the combination of direct measurements, indirect
measurements, and theoretical inputs.
It is hoped that this paper will be helpful in such efforts.

This paper is organized as follows. First, in Secs.~\ref{sec:single_particle}
and~\ref{sec:further}, the concepts of single-particle
wave functions and reduced-width amplitudes are introduced. This discussion
includes several methods of defining resonances energies and widths as well as
computation methods.
Most of the calculations in this paper utilize $R$-matrix theory,
with the tail of the nuclear potential beyond the channel radii included.
This approach allows single-particle quantities, such
as widths and ANCs, to be calculated in the typical manner
from Woods-Saxon potentials. At the same time, these quantities can
be described in the $R$-matrix framework using single-particle reduced-width
amplitudes and penetration factors.
The inclusion of multiple channels is also straightforward in the
$R$-matrix approach.
Next, in Sec.~\ref{sec:overlaps}, the concepts of spectroscopic factors
and overlaps are introduced using the same language and connected to
$R$-matrix theory.
Then, in Sec.~\ref{sec:isospin}, these concepts are applied to isospin
and mirror symmetry.
Mirror symmetry is then investigated in Sec.~\ref{sec:single} using the
examples of nucleon~+~${}^{12}{\rm C}$, nucleon~+~${}^{16}{\rm O}$,
and nucleon~+~${}^{26}{\rm Al}$.
All of these examples involve $\ell=0$ nucleons and energy levels near
the nucleon separation threshold, where the effects of continuum coupling
may be significant. Several different approaches are compared.
Finally, in Sec.~\ref{sec:multi}, the effect of
mirror symmetry operating on a set of levels is considered.
In this situation, there is a mixing of the levels due to mirror symmetry
breaking beyond the channel radii.
The $R$-matrix approach presented here is a new and efficient
method for investigating this question.
These effects are demonstrated using the first three $2^+$ states of
${}^{18}{\rm O}$ and ${}^{18}{\rm Ne}$.
Appendix~\ref{app:dlde} describes an algorithm that is useful for determining
the contribution of the wave-function tail to the overall normalization in
a Coulomb potential.

\section{Single-particle wave functions}
\label{sec:single_particle}

Single-particle wave functions provide a basis for the approximate
description of the many-body nuclear physics problem.
While the term ``single-particle'' is appropriate for the nucleon~+~nucleus
case, one can consider these wave functions more generally as one-body
wave functions describing a particular two-cluster configuration of a many-body
nuclear wave function.
The single-particle radial wave function $u(r)/r$ is assumed to satisfy
the radial Schr{\" o}dinger equation
\begin{equation} \label{eq:schrodinger}
-\frac{\hbar^2}{2\mu}\frac{d^2u}{dr^2} + [V(r)+V_C(r)]u
+\frac{\hbar^2}{2\mu}\frac{\ell(\ell+1)}{r^2}u=Eu,
\end{equation}
where $r\ge 0$ is the distance between the clusters,
$E$ is the relative energy,
$\ell$ is the relative orbital angular momentum quantum number,
$\mu$ is the reduced mass, and $\hbar$ is Planck's constant.
The nuclear single-particle potential $V(r)$ is assumed to be real, central,
and local, with $\lim_{r\rightarrow 0} r^2V(r)=0$ and
$\lim_{r\rightarrow\infty} r^2V(r)=0$, and $V_C(r)$ is the Coulomb potential.
Physical solutions will have $u(r) \propto r^{\ell+1}$ for $r\rightarrow 0$,
which provides a boundary conditions for $u(0)$.
This single-particle wave function is specific to a particular channel,
where a channel is defined to be a configuration of given cluster type,
total angular momentum, parity, orbital angular momentum, and channel spin.
This approach ignores any coupling between different channels.

When the energy $E$ corresponds to a bound or unbound energy level,
the single-particle reduced-width amplitude $\gamma$ is defined by
\begin{equation} \label{eq:reduced_width}
  \gamma=u(a) \left(\frac{\hbar^2}{2\mu a \int_0^a u^2\,dr}\right)^{1/2}.
\end{equation}
The reduced-width amplitude has the physical interpretation of being the
amplitude of the resonant wave function at the channel radius, when the
wave function is normalized to unity inside the channel radius.
All of the ANCs, widths, and reduced widths in this section are derived
from single-particle wave functions and are single-particle quantities.

In what follows, I will use $r$ (possibly with a subscript)
to indicate an arbitrary radius,
$b$ to indicate a large radius where $V(r)$ is negligible, and
$a$ to indicate a channel radius, which is located outside the nuclear
surface where $V(r)$ is small but not necessarily negligible.
In  this section, I define the Coulomb functions used in this work
and discuss two integral relations that are useful for single-particle states.
Then three slightly different ways of defining resonances are introduced and
discussed.

\subsection{Coulomb functions}
\label{subsec:coulomb}

In regions where only the point-Coulomb potential is present, the solutions
to Eq.~(\ref{eq:schrodinger}) are given by
$u=G_\ell(\eta,\rho)\equiv G$ and $u=F_\ell(\eta,\rho)\equiv F$, which are
the irregular and regular Coulomb functions, respectively.
I also define $\rho=kr$, $k=\sqrt{2\mu E/\hbar^2}$, and
$\eta k = Z_1Z_2q^2\mu/\hbar^2$, where $Z_1q$ and $Z_2q$ are the charges of
the two clusters.
The Wronskian relation for the Coulomb functions is
\begin{equation} \label{eq:wronskian}
G\frac{dF}{dr}-F\frac{dG}{dr}=k .
\end{equation}
Outgoing and incoming Coulomb waves are defined via
\begin{subequations}
\begin{eqnarray}
 O &=& \exp(-i\sigma)(G + iF) \quad\quad {\rm and} \\
 I &=& \exp(i\sigma)(G - iF),
\end{eqnarray}
\end{subequations}
where $\sigma(\ell,\eta)$ is the Coulomb phase shift.
One also has
\begin{equation}
O = \exp\left[\frac{\pi}{2}(\eta-i\ell)\right]W_{-i\eta,\ell+1/2}(-2i\rho),
\end{equation}
where $W$ is the Whittaker function.
For $E$ real and negative, such as is the case for bound states,
I take $k=i\sqrt{-2\mu E/\hbar^2}$, and $W$ is real.
I also consider situations where $E$ is complex, with $\operatorname{Re}{E}>0$
and  $\operatorname{Im}{E}<0$, in which case the sign of
$k=\sqrt{2\mu E/\hbar^2}$ defined such that  $\operatorname{Re}{k}>0$ and
$k$ is located near the physical (i.e., real and positive) $k$~axis.
The logarithmic derivative of the outgoing solution by
\begin{equation} \label{eq:L_outgoing}
L \equiv \frac{r}{O}\frac{dO}{dr},
\end{equation}
and when $E$ is real one also defines
\begin{equation}
L \equiv \hat{S}+iP,
\end{equation}
where $\hat{S}$ and $P$ are the shift and penetration factors, respectively.
Note that $P$ vanishes for $E\le 0$.
Finally, the phase $\phi$ is defined by $\tan\phi=F/G$.
The functions $F$, $G$, $O$, $I$, $W$, $L$, $\hat{S}$, $P$, and $\phi$
are useful for large radii, where $V(r)$ is negligible.
They may be continued to smaller
radii using the differential equation, Eq.~(\ref{eq:schrodinger}), to
yield the nuclear-modified Coulomb functions $\mathcal{F}$, $\mathcal{G}$,
$\mathcal{O}$, $\mathcal{I}$, $\mathcal{W}$, $\mathcal{L}$,
$\hat{\mathcal{S}}$, $\mathcal{P}$, and $\Phi$.
Where applicable, these modified Coulomb functions obey the
same Wronskian relations as the usual Coulomb functions because of the
differential equation they satisfy, Eq.~(\ref{eq:schrodinger}).

\subsection{Two integral relations}

Here I derive two integral relations which enable the extraction of
ANCs or widths from the single-particle radial wave function.
The first integral relation concerns the energy derivative of the
logarithmic radial derivative of $u$. An early reference for this
procedure is \textcite[V.1, Eqs. (1.5)-(1.9), p.~283]{Lan58}.
Using Eq.~(\ref{eq:schrodinger}) with two different solutions
$u_1$ and $u_2$ corresponding to energies $E_1$ and $E_2$, one can show that
\begin{equation} \label{eq:int_o1o2}
-\frac{\hbar^2}{2\mu}\frac{d}{dr}\left[u_1\frac{du_2}{dr}-u_2\frac{du_1}{dr}
  \right]=(E_2-E_1) u_1u_2.
\end{equation}
Upon integrating from $r=r_1$ to $r_2$, with $r_1<r_2$, this becomes
\begin{equation}
\begin{split}
  -\frac{\hbar^2}{2\mu}\left[u_1u_2 
  \left(\frac{1}{u_2}\frac{du_2}{dr} \right.\right. & \left.\left.
  -\frac{1}{u_2}\frac{du_1}{dr}\right)
  \right]_{r_1}^{r_2} \\ & = (E_2-E_1)\int_{r_1}^{r_2} u_1u_2 \, dr.
\end{split}
\end{equation}
Taking $u_1\rightarrow u_2$ yields
\begin{equation}
-\frac{\hbar^2}{2\mu}\left[ u^2\frac{\partial}{\partial E}\left(\frac{1}{u}
  \frac{du}{dr}\right)\right]_{r_1}^{r_2} =\int_{r_1}^{r_2} u^2 \, dr ,
\label{eq:dlogude}
\end{equation}
where $\partial E$ is taken at fixed radius.
Using the boundary condition on $u(0)$, one can take $r_1\rightarrow 0$
to obtain
\begin{equation} \label{eq:dlogude_zero}
-\frac{\hbar^2}{2\mu}u^2(r_2)\left[\frac{\partial}{\partial E}\left(\frac{1}{u}
  \frac{du}{dr}\right)\right]_{r_2} =\int_0^{r_2} u^2 \, dr .
\end{equation}

The second relation is an application of the two-potential
formalism~\cite[X.V.17, pp.~404-405]{Mes61}.
The regular Coulomb wave function $F$ is the solution to
Eq.~(\ref{eq:schrodinger}) for the point-Coulomb potential alone,
$V_{pC}=Z_1Z_2q^2/r$, while $u$ is the solution for $V+V_C$.
Taking the differential equation satisfied by $u$ multiplied by
$F$ and subtracting the differential equation satisfied by $F$ multiplied by
$u$, I obtain
\begin{equation}
\frac{\hbar^2}{2\mu}\frac{d}{dr}\left(F\frac{du}{dr}-u\frac{dF}{dr}\right) =
  F(V+V_C-V_{pC})\, u.
\end{equation}
Upon integrating from $r=r_1$ to $r_2$, with $r_1<r_2$, this becomes
\begin{equation}
\frac{\hbar^2}{2\mu}\left(F\frac{du}{dr}-u\frac{dF}{dr}\right)_{r_1}^{r_2} =
\int_{r_1}^{r_2} F(V+V_C-V_{pC})\, u \, dr.
\end{equation}
I now specify $r_1\rightarrow 0$ and $r_2 =b$, where $V+V_C-V_{pC}$
becomes negligible, and $u(r)=\alpha F(r) + \beta G(r)$ in the vicinity
of $r=b$, where $\alpha$ and $\beta$ are constants.
Using the Wronskian relation, Eq.~(\ref{eq:wronskian}), this becomes
\begin{equation} \label{eq:integral_asymp}
-\beta\frac{\hbar^2 k}{2\mu}=\int_0^b F(V+V_C-V_{pC})\, u \, dr.
\end{equation}

\subsection{{\em S} Matrix}

The radial wave function may be written as a linear combination
of modified Coulomb functions
\begin{equation} \label{eq:s_matrix}
u \propto \mathcal{I}-S\mathcal{O},
\end{equation}
where $S$ is the scattering matrix. This relation may also be expressed as
\begin{equation} \label{eq:phase_shift}
u \propto \cos\delta \,\mathcal{F} + \sin\delta \,\mathcal{G},
\end{equation}
where $\delta$ is the phase shift and
\begin{equation}
S=\exp[2i(\delta+\sigma)].
\end{equation}
It is important to note that Eqs.~(\ref{eq:s_matrix})
and~(\ref{eq:phase_shift}) are valid for any radius, although they will
only be used for $r\ge a$.

In the $S$-matrix approach, discrete energy levels may be defined to be
those energies where the solution consists of a pure outgoing wave.
This provides the usual large-radius boundary condition for bound states,
where $E$ is real and negative.
For unbound states, this boundary condition can only
be achieved for complex $E$.
If there is such an energy level at $E=E_0$, then the $S$~matrix has a
first-order pole at the energy such that near $E=E_0$
\begin{equation}
S(E)=\frac{A}{E-E_0}+\mbox{function that is regular at $E_0$},
\end{equation}
where $A$ is the residue.
Using Eq.~(\ref{eq:s_matrix}), one finds
\begin{equation}
\frac{1}{u}\frac{du}{dr}=
  \frac{S^{-1}\frac{d\mathcal{I}}{dr}+\frac{d\mathcal{O}}{dr}}%
  {S^{-1}\mathcal{I}+\mathcal{O}} .
\end{equation}
The energy derivative of the logarithmic radial derivative of $u$ at $E=E_0$
may then be evaluated by substituting this result into
Eq.~(\ref{eq:dlogude_zero}). By using the Wronskian relation
\begin{equation}
\mathcal{I}\frac{d\mathcal{O}}{dr}-\mathcal{O}\frac{d\mathcal{I}}{dr}=2ik,
\end{equation}
one obtains
\begin{equation} \label{eq:inf_s_norm}
\begin{split}
-\frac{2ik}{A}\frac{u^2(r_2)}{\mathcal{O}^2(r_2)}=\frac{2\mu}{\hbar^2}
  & \left[ \int_0^{r_2} u^2\, dr \right. \\ & \left.
  + \frac{\hbar^2}{2\mu r_2}u^2(r_2)
  \left(\frac{\partial\mathcal{L}}{\partial E}\right)_{r_2}\right] .
\end{split}
\end{equation}
This equation is independent of $r_2$, since on the left side one has
$u(r)\propto\mathcal{O}(r)$ by definition and Eq.~(\ref{eq:dlogude})
shows the right side is independent of $r_2$.

If the level is bound, one may normalize the $u$ over all space by
requiring that the quantity in brackets on the right side of
Eq.~(\ref{eq:inf_s_norm}) is equal to unity. This result can be seen
by taking $r_2 \rightarrow\infty$ where the surface term vanishes and
the usual bound-state normalization condition emerges.
It is also shown by \textcite[Eq.~(A.29), p.~351]{Lan58}.
However, since it is convenient to consider different normalizations,
I will leave the normalization unspecified and instead define
\begin{equation} \label{eq:I_infinity}
I_\infty =  \int_0^{r_2} u^2\, dr + \frac{\hbar^2}{2\mu r_2}u^2(r_2)
  \left(\frac{\partial\mathcal{L}}{\partial E}\right)_{r_2}.
\end{equation}
One also has
\begin{equation} \label{eq:ANC_1}
\frac{u(r)}{I_\infty^{1/2}} = C\mathcal{W}(r) =
  C \exp\left[-\frac{\pi}{2}(\eta-i\ell)\right]\,\mathcal{O}(r) ,
\end{equation}
where $C$ is the ANC, which is a real quantity.
There is also a simple relationship between the ANC and the residue:
\begin{equation} \label{eq:ANC_2}
C^2 = i\exp[\pi(\eta-i\ell)] \, \frac{\mu A}{\hbar^2 k} .
\end{equation}

If the level is unbound, the all-space normalization may also be achieved
by normalizing $u$ such that $I_\infty=1$.
In this case, the normalization is less obvious, because the
integral is not convergent in the usual sense as $r_2 \rightarrow\infty$.
However, this regularization procedure has been shown to be useful and also
consistent with the Zel'dovich regularization method which involves inserting
a convergence factor into the integrand~\cite{Zel61,Gya71,Gar76}.
Since the level energy $E_0$ is complex in this case, one may define
the real and imaginary parts according to
\begin{equation}
E_0 = E_S - i\frac{\Gamma_S}{2},
\end{equation}
where $E_S$ is the resonance energy defined by the $S$-matrix pole
and $\Gamma_S$ is the corresponding width.
In addition, the pole residue can be used to define a width via
\begin{equation} \label{eq:Gamma_S1}
\Gamma_{S1} = |A| .
\end{equation}
In general, $\Gamma_S\ne\Gamma_{S1}$, but they become equal in the limit
$\Gamma_S\ll E_S$.

Another expression for $\Gamma_S$ may be found by multiplying
Eq.~(\ref{eq:schrodinger}) by $u^*$, subtracting the complex
conjugate, and then integrating~\cite{Hum61}:
\begin{subequations} \label{eq:gamma_current}
\begin{align}
\Gamma_S &= i\frac{\hbar^2}{2\mu}\frac{\left(u\frac{du^*}{dr}-u^*\frac{du}{dr}
  \right)_{r_2}}{\rule{0pt}{1.0em} \int_0^{r_2} |u|^2 \, dr} \\
  &= i\frac{\hbar^2}{2\mu r_2}\frac{ |u(r_2)|^2
  (\mathcal{L}^*-\mathcal{L})_{r_2}}%
  {\rule{0pt}{1.0em} \int_0^{r_2} |u|^2 \, dr} .
\end{align}
\end{subequations}
This formula is useful when $\Gamma_S$ is very small and other approaches
to calculating $\Gamma_S$ may not be accurate~\cite{Kru04}.
In this situation, one can use the first-order Taylor series
\begin{equation}
\mathcal{L}\approx \hat{\mathcal{S}}+i\mathcal{P}-i\frac{\Gamma_S}{2}
  \left( \frac{\partial\hat{\mathcal{S}}}{\partial E}+
  i\frac{\partial\mathcal{P}}{\partial E} \right),
\end{equation}
where $\hat{\mathcal{S}}$, $\mathcal{P}$, and their energy derivatives are
evaluated on the real energy axis at $E_S$ and are real quantities.
Defining for convenience an alternative reduced width amplitude
\begin{equation}
|\bar{\gamma}|^2 = \frac{\hbar^2}{2\mu r_2}
  \frac{|u(r_2)|^2}{\rule{0pt}{1.0em} \int_0^{r_2} |u|^2 \, dr } ,
\end{equation}
yields
\begin{equation} \label{eq:gamma_current_approx}
\Gamma_S \approx \frac{2|\bar{\gamma}|^2\mathcal{P}(r_2)}%
  {1+|\bar{\gamma}|^2
  \left(\frac{\partial\hat{\mathcal{S}}}{\partial E}\right)_{r_2}} ,
\end{equation}
a formula that is very accurate when $\Gamma_S\ll E_S$.
Equations~(\ref{eq:gamma_current}) and~(\ref{eq:gamma_current_approx}) are
valid for all $r_2$, but are most easily evaluated for $r_2=b$.

One also has
\begin{equation}
\left[\frac{u(b)}{O(b)}\right]^2 = i\frac{\mu A I_\infty}{\hbar^2 k}
\end{equation}
for large radii.
The integral relation given by Eq.~(\ref{eq:integral_asymp}) yields
\begin{equation} \label{eq:asymp_integral}
\frac{u(b)}{O(b)} = -\frac{2\mu}{\hbar^2k}\exp(i\sigma)
 \int_0^b F(V+V_C-V_{pC})\, u \, dr ,
\end{equation}
which provides an alternative method of calculating the residue
and hence also the ANC or $\Gamma_{S1}$.
Finally, by adopting $r_2=a$ in Eq.~(\ref{eq:inf_s_norm}), the residue
may be expressed in terms of the reduced-width amplitude:
\begin{equation} \label{eq:amplitude_reduced}
A = -2i \, \frac{ka}{\mathcal{O}^2(a)} \,
  \frac{\gamma^2}%
  {1+\gamma^2\left(\frac{\partial\mathcal{L}}{\partial E}\right)_a} .
\end{equation}

\subsection{{\em R} Matrix}
\label{subsec:r_matrix}

In the $R$-matrix approach, the logarithmic derivative of $u$ at the
channel radius $a$ is described by the $R$~matrix, which is a scalar
quantity in the single-channel case.
Specifically~\cite[IV.2, Eq.~(2.4), p.~274]{Lan58},
\begin{equation} \label{eq:R_log_deriv}
\left(\frac{r}{u}\frac{du}{dr}\right)_a=R^{-1}+B,
\end{equation}
where $B$ is the real boundary condition constant.
$R$-matrix energy levels are defined by the poles of the $R$~matrix,
where the logarithmic derivative of $u$ is $B/a$.
The eigenfunctions satisfying this boundary condition form a complete set
inside the channel radius
and it can be shown~\cite[IV, pp.~272-274]{Lan58} that
\begin{equation}
R = \sum_\lambda \frac{\gamma_\lambda^2}{E_\lambda-E},
\end{equation}
where $\gamma_\lambda$ are the reduced width amplitudes and
$E_\lambda$ are the level energies.
Note that these reduced widths, defined as residues of $R$-matrix poles,
are completely consistent with the definition given by
Eq.~(\ref{eq:reduced_width}) and Eqs.~(\ref{eq:dlogude_zero})
and~(\ref{eq:R_log_deriv}).
In order to investigate a level at an energy $E_R$ in the $R$-matrix
approach, it is natural to choose $B=\hat{\mathcal{S}}(E_R)$,
the real part of the outgoing wave boundary condition.

If the level is bound, this boundary condition is unchanged from the
$S$-matrix case and the results for $u$ are identical.
Rewriting Eq.~(\ref{eq:inf_s_norm}) with $r_2=a$, taking into account the
definition of the reduced width amplitude, replacing the residue and
$\mathcal{O}$ with the appropriately normalized ANC and $\mathcal{W}$,
and noting
$\partial\mathcal{L}/\partial E=\partial\hat{\mathcal{S}}/\partial E$
for bound states yield
\begin{equation} \label{eq:R_matrix_ANC}
C^2 = \frac{2\mu a}{\hbar^2 \mathcal{W}^2(a)} \frac{\gamma^2}%
  {\rule{0pt}{1.0em} 1+
  \gamma^2\left(\frac{\partial\hat{\mathcal{S}}}{\partial E}\right)_a} .
\end{equation}

If the level is unbound, the boundary condition implies that
\begin{equation}
u(r)=u(a)\frac{\mathcal{F}(a)\mathcal{F}(r)+\mathcal{G}(a)\mathcal{G}(r)}%
  {\mathcal{F}^2(a)+\mathcal{G}^2(a)}.
\end{equation}
Unlike the other cases, this condition depends somewhat on the value of $a$.
The $R$-matrix expression for the phase shift is
\begin{equation}
\delta = -\Phi(a) +
  \tan^{-1}\frac{\mathcal{P}(a)}{R^{-1}-\hat{\mathcal{S}}(a)+B}.
\end{equation}
A convenient definition of the width $\Gamma_R$ is provided by
\begin{equation}
\left(\frac{d\delta}{dE}\right)_{E_R} =
  -\left(\frac{d\Phi}{dE}\right)_{E_R}+\frac{2}{\Gamma_R},
\end{equation}
that implies
\begin{equation} \label{eq:R_matrix_width}
\Gamma_R=\frac{2\gamma^2\mathcal{P}(a)}%
  {\rule{0pt}{1.0em} 1+
  \gamma^2\left(\frac{\partial\hat{\mathcal{S}}}{\partial E}\right)_a} .
\end{equation}
Note that this expression for the width is very similar in structure
to Eq.~(\ref{eq:R_matrix_ANC}), the equation for the ANC.
For unbound states, the integral relation given by
Eq.~(\ref{eq:integral_asymp}) yields
\begin{equation} \label{eq:r_int_asymp}
\frac{u(a)\mathcal{G}(a)}{\mathcal{F}^2(a)+\mathcal{G}^2(a)} =
-\frac{2\mu}{\hbar^2 k}  \int_0^b F(V+V_C-V_{pC})\, u \, dr ,
\end{equation}
which provides another way to calculate the reduced-width amplitude.
Note that unbound states in the $R$-matrix approach cannot be normalized
to unity over all space via the regularization procedure that may be applied
for the case of Gamow states.
The normalization is left unspecified, although it is often
assumed that $\int_0^a u^2 \, dr=1$.

\subsection{{\em K} matrix}

Here I will consider an unbound state with a $K$-matrix boundary condition,
where the resonance energy is defined to have a phase shift
of $\delta=\pi/2+m\pi$, where $m$ is an integer and
$u(r)\propto\mathcal{G}(r)$ at the resonance energy (see
\textcite[Eq.~9.4]{Hum90}). Using
\begin{equation}
\frac{1}{u}\frac{du}{dr}=\frac{\cos\delta\,\frac{d\mathcal{F}}{dr}+
  \sin\delta\,\frac{d\mathcal{G}}{dr}}%
  {\cos\delta\,\mathcal{F}+\sin\delta\,\mathcal{G}},
\end{equation}
one finds at a $K$-matrix resonance
\begin{equation} \label{eq:K_matrix_deriv}
\begin{split}
\left[\frac{\partial}{\partial E} \right. & \left.
  \left(\frac{r}{u}\frac{du}{dr} \right)\right]_{r_2} \\
  & =\left[\frac{\partial}{\partial E} \left(\frac{r}{\mathcal{G}}
  \frac{d\mathcal{G}}{dr}
  \right)\right]_{r_2} -\frac{kr_2}{\mathcal{G}^2(r_2)}\frac{d\delta}{dE} .
\end{split}
\end{equation}
The $K$-matrix width $\Gamma_K$ may be defined at the $K$-matrix resonance
energy $E_K$ via
\begin{equation}
\left(\frac{d\delta}{dE}\right)_{E_K} = \frac{2}{\Gamma_K}.
\end{equation}
Then using Eqs.~(\ref{eq:dlogude_zero}) and (\ref{eq:K_matrix_deriv})
one finds
\begin{equation}
\begin{split}
\Gamma_K = & \frac{\hbar^2 k}{\mu}\frac{u^2(r_2)}{\mathcal{G}^2(r_2)}
  \left\{ \int_0^{r_2} u^2 \, dr\right. \\ & \left.
  +\frac{\hbar^2}{2\mu r_2} u^2(r_2)
  \left[\frac{\partial}{\partial E}\left(\frac{r}{\mathcal{G}}
  \frac{d\mathcal{G}}{dr}\right)\right]_{r_2}\right\}^{-1} .
\end{split}
\end{equation}
The integral relation given by Eq.~(\ref{eq:integral_asymp}) yields
\begin{equation}
\frac{u(b)}{G(b)} =
-\frac{2\mu}{\hbar^2k}  \int_0^b F(V+V_C-V_{pC})\, u \, dr .
\end{equation}

\subsection{Practical calculations}
\label{subsec:practical}

For calculations, I assume that $V(r)$ is described by a
phenomenological Woods-Saxon potential
\begin{equation}
V(r) = \frac{-V_0}{1+\exp[(r-R_n)/a_n]}
\end{equation}
and the Coulomb potential is given by a uniformly-charged sphere of
radius $R_C$.
In this work, numerical calculations will only be performed for
nucleon~+~$A$-nucleon configurations, although all equations are
completely general and apply, for example, to $\alpha$~+~nucleus channels.
It is assumed that $R_n=r_nA^{1/3}$ and
$R_C=r_CA^{1/3}$, with $r_n=r_C=1.25$~fm
and a diffuseness parameter of $a_n=0.65$~fm, unless otherwise specified.
The potential depth $V_0$ is often adjusted to reproduce the level energy and
$n$, the number of radial nodes inside the channel radius (including
the origin).
When $E$ is complex, $u(r)$ is likewise complex, and the nodes are counted
using the real part of $u(r)$, where the phase of $u(r)$ is fixed such that
$u(r)$ is real and positive near $r=0$.

Equation~(\ref{eq:schrodinger}) is solved numerically via the Numerov method.
In the case of $u$, either the potential depth is varied to generate the
desired level energy, or the level energy is determined for a fixed
potential. One solution is propagated outward
from $r=0$, starting with $u(0)=0$. Another solution
is propagated inward from $r=b$, starting with the desired boundary
condition. The solutions are compared at the channel radius $r=a$.
The level energy or potential depth is then varied to reproduce the
desired $n$ value and match the logarithmic derivatives at the channel radius.
The modified Coulomb functions are also found by numerical integration,
starting with the unmodified Coulomb functions at $r=b$ and integrating
inward to $r=a$. I utilize $a=R_n+a_n$, unless otherwise specified,
and $b=20$~fm.

In the preceding development, energy derivatives of $\mathcal{L}$,
$\hat{\mathcal{S}}$, $\mathcal{P}$, and $(r/\mathcal{G})(d\mathcal{G}/dr)$
play an important role.
For $r=b$, where the Coulomb functions are unmodified, $\partial L/\partial E$
can be efficiently calculated using the continued fraction algorithm given
in Appendix~\ref{app:dlde}.
This algorithm is of general use for the normalization of bound and
Gamow states, as well as for $R$-matrix calculations.
For smaller radii, one has from Eq.~(\ref{eq:dlogude})
\begin{equation}
-\frac{\hbar^2}{2\mu}\left[ \frac{\mathcal{O}^2}{r}
  \left(\frac{\partial\mathcal{L}}{\partial E}\right)\right]_{r_1}^b
  =\int_{r_1}^b \mathcal{O}^2 \, dr
\end{equation}
that allows $\partial\mathcal{L}/\partial E$ at smaller radii to be calculated
from $\partial L/\partial E$ at $r=b$ and an integration.
Since $\mathcal{L}=\hat{\mathcal{S}}+i\mathcal{P}$, this method takes care
of three out of four of the needed energy derivatives.
Alternatively, $\partial\mathcal{L}/\partial E$ may be calculated by
numerical differentiation, an approach that also works for
the remaining case, the energy derivative of
$(r/\mathcal{G})(d\mathcal{G}/dr)$, that in practice is only needed for $r=b$.

\subsection{Discussion}
\label{subsec:single_discuss}

\begin{table*}[tbh]
\caption{Effect of the resonance definition on the resonance energy
  ($E_S$, $E_R$, and $E_K$) and resonance width ($\Gamma_S$, $\Gamma_{s1}$,
  $\Gamma_R$, and $\Gamma_K$), for three situations.}
\label{tab:resonance_def}
\begin{tabular}{ccccccccccc} \hline\hline \\[-2.0ex]
system & $n\ell$ & $E_S$ & $E_R$ & $E_K$ &
  \multirow{2}{*}{$\frac{E_S-E_R}{\Gamma_S}$} &
  \multirow{2}{*}{$\frac{E_S-E_K}{\Gamma_S}$} &
  $\Gamma_S$ & $\Gamma_{S1}$ & $\Gamma_R$ & $\Gamma_K$ \\
&      & (keV) & (keV) & (keV) &&&
  (keV) & (keV) & (keV) & (keV) \\ \hline \\[-2.0ex]
$p+{}^{12}{\rm C}$  & $2s$ & 420.5 & 424.0 & 424.1 & $-0.0921$ & $-0.0935$ &
  37.50 & 35.85 & 39.17 & 39.23 \\
$p+{}^{14}{\rm N}$  & $2s$ & 259.3 & 259.3 & 259.3 & $-0.0072$ & $-0.0072$ &
  1.173 & 1.169 & 1.173 & 1.173 \\
$p+{}^{26}{\rm Al}$ & $2s$ & 126.8 & 126.8 & 126.8 & -         & -         &
  $4.96\times 10^{-9}$ & $4.96\times 10^{-9}$ & $4.96\times 10^{-9}$ &
  $4.96\times 10^{-9}$ \\
\hline\hline
\end{tabular}
\end{table*}

\begin{figure}
\includegraphics[width=\columnwidth]{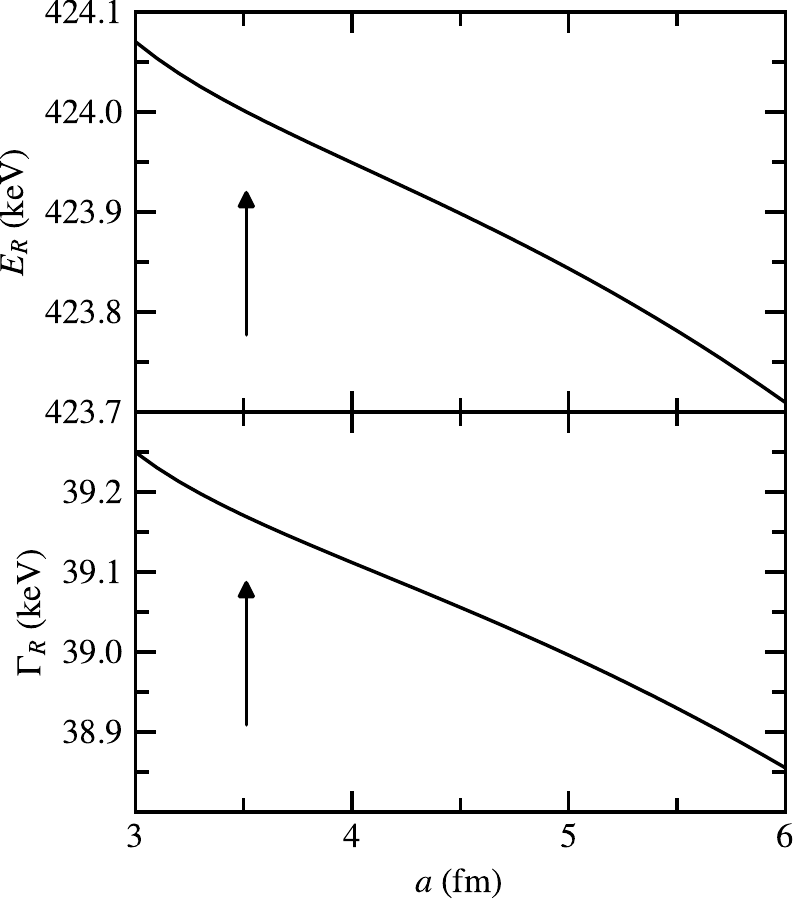}
\caption{The dependence of the $R$-matrix $p+{}^{12}{\rm C}$ resonance energy
and width on the channel radius. The arrows indicate the channel radius
of 3.51~fm used for the calculations reported in
Table~\ref{tab:resonance_def}.} \label{fig:c12_p_chan_rad}
\end{figure}

The different resonance definitions do give rise to somewhat different
resonance energies and widths when the resonances become broad.
This effect has been investigated by considering $\ell=0$ resonances in
$p+{}^{12}{\rm C}$, $p+{}^{14}{\rm N}$, and $p+{}^{26}{\rm Al}$.
The potential depth was adjusted once to reproduce the resonance energy
for $R$-matrix definition, and was then left constant for the calculation
with the other definitions.
The results are shown in Table~\ref{tab:resonance_def}, where
the resonance energy for $p+{}^{12}{\rm C}$ is from Ref.~\cite{Mey76},
that for $p+{}^{14}{\rm N}$ is from Ref.~\cite{Dai16}, and that for
$p+{}^{26}{\rm Al}$ is from Refs.~\cite{Lot11,Wan17}.
Because of the variable Coulomb barrier and resonance energy, the
single-particle width vary considerably.
In the case of $p+{}^{12}{\rm C}$, the single-particle width is just under
10\% of the resonance energy and non-negligible differences in the
resonance energies and widths are seen.
The differences in resonance energy are seen to be a small fraction
of the resonance width.

These differences are easily understood. For example, the single-channel
$S$ and $R$~matrices are related via
\begin{equation}
S = \frac{2i\rho}{\mathcal{O}^2} ( R^{-1}-\hat{\mathcal{S}}-i\mathcal{P}+B )^{-1}
  + \frac{\mathcal{I}}{\mathcal{O}} .
\end{equation}
The energy dependence of the shift and penetration factors may be
approximated using
\begin{equation}
\hat{\mathcal{S}}+i\mathcal{P} \approx B+i\mathcal{P}_R +
  (E-E_R)\left( \frac{\partial\hat{\mathcal{S}}}{\partial E}+
  i\frac{\partial\mathcal{P}}{\partial E} \right)_{E_R},
\end{equation}
where $\mathcal{P}_R$ and the energy derivatives are evaluated at $E_R$.
Making a single-level approximation to $R$ then allows the $S$- and $R$-matrix
pole positions to be related:
\begin{equation} \label{eq:S_from_R}
E_S-i\frac{\Gamma_S}{2} \approx E_R -i \frac{\gamma^2 \mathcal{P}_R}%
  {1+\gamma^2\left(
  \frac{\partial\hat{\mathcal{S}}}{\partial E}+
  i\frac{\partial\mathcal{P}}{\partial E} \right)_{E_R} } ,
\end{equation}
where $\gamma$ is the $R$-matrix reduced width.

The $R$-matrix resonance energy and width also have some dependence
on the value of the channel radius.
This variation is shown for the case of the $p+{}^{12}{\rm C}$ resonance
and $3\le a\le 6$~fm in Fig.~\ref{fig:c12_p_chan_rad}, where both
$E_R$ and $\Gamma_R$ are seen to vary by about 0.4~keV.
However, the $E_S$ and $\Gamma_S$ calculated from the $R$-matrix pole
parameters using Eq.~(\ref{eq:S_from_R}) only vary within
0.12 and 0.05~keV, respectively.

The differences between the various resonance energy and width definitions
are small, unless the width is not small compared to the resonance energy.
Different resonance energy definitions are also discussed in Ref.~\cite{For06},
where similar conclusions are reached.
These approaches should be viewed as different, but equivalent, descriptions
of the same resonance.
In practice, if the choice matters, it should be dictated by consistency with
how the single-particle state is used.
If a transfer reaction or $R$-matrix calculation is coupled
to a single-particle calculation, consistent resonance
definitions should be utilized throughout.
For example, some versions of the transfer reaction code
{\sc dwuck4}~\cite{Kun08} utilizes
the $K$-matrix boundary condition to define a resonant state.
In addition, if the resonance is not narrow, it is unlikely to be a
good approximation to treat it in isolation. Rather, the effects of
potential scattering and/or interference with other resonances
will be significant.

Although the $S$- and $K$-matrix approaches are formally independent of
the channel radius, all three approaches can also be unified from
a generalized $R$-matrix point of view with a channel radius~\cite{Lan66}.
See also \textcite{Kap38} for the $S$-matrix pole expansion using
a channel radius. In the $K$-matrix case, the results can be recast into
$R$-matrix form by noting the $K$-matrix boundary condition corresponds to
adopting $B=(r/\mathcal{G})(d\mathcal{G}/dr)$ at the channel radius.
For the remainder of this work, I will utilize the $R$-matrix definition
of resonances parameters almost exclusively.

It should also be kept in mind that these results depend to various
degrees on the single-particle potential parameters and channel radius.
The reduced widths vary with radius of the
nuclear potential (which depends on $A$), the orbital angular momentum,
and number of radial nodes. The systematics of these variations have been
studied by \textcite{Ili97}, where substantial variations are seen, even
if the reduced width is made dimensionless.
The variation of the single-particle reduced width with charge and energy
is weaker, provided the energy variation stays within a few MeV of
the separation threshold. The reason for this observation is that Coulomb
energy difference or potential energy change is relatively small compared to
the depth of the nuclear potential, which accordingly leads to a
small change in the wave function inside the channel radius.

\section{Further development}
\label{sec:further}

\subsection{Effect of the nuclear potential on the penetrability}
\label{subsec:nuc_pen}

\begin{figure}
\includegraphics[width=\columnwidth]{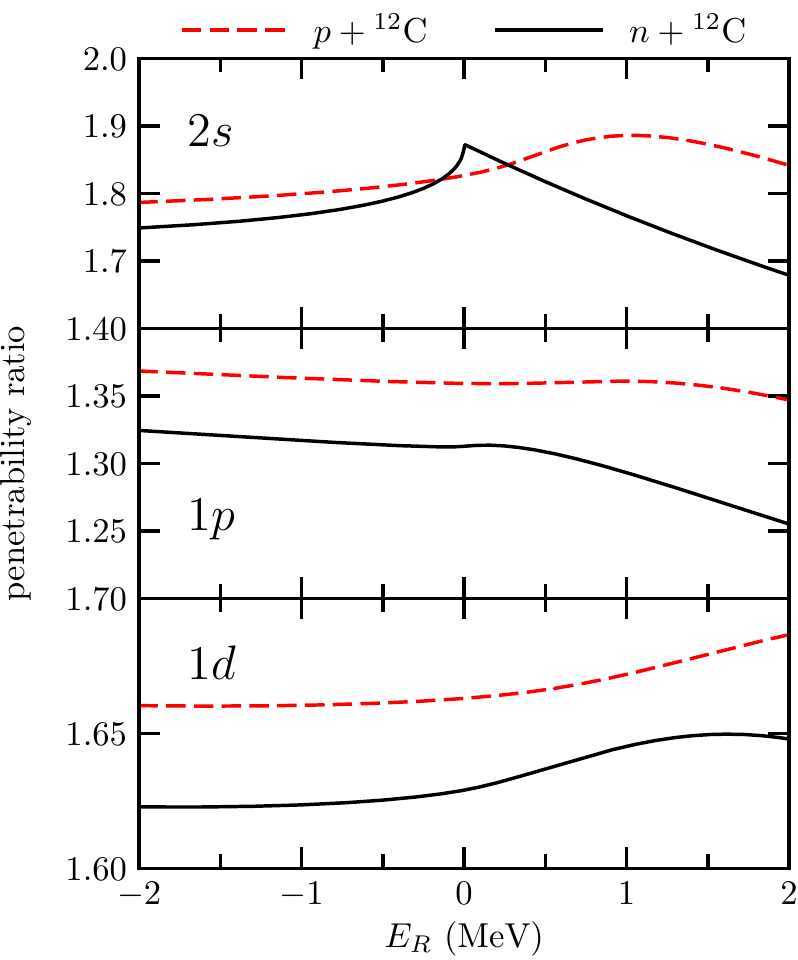}
\caption{The penetrability ratio as a function of the level energy
for $p+{}^{12}{\rm C}$ and $n+{}^{12}{\rm C}$ for $n\ell=2s$, $1p$, and $1d$.
See Subsec.~\ref{subsec:nuc_pen} for details.} \label{fig:pen_ratio_c12_N}
\end{figure}

\begin{figure}
\includegraphics[width=\columnwidth]{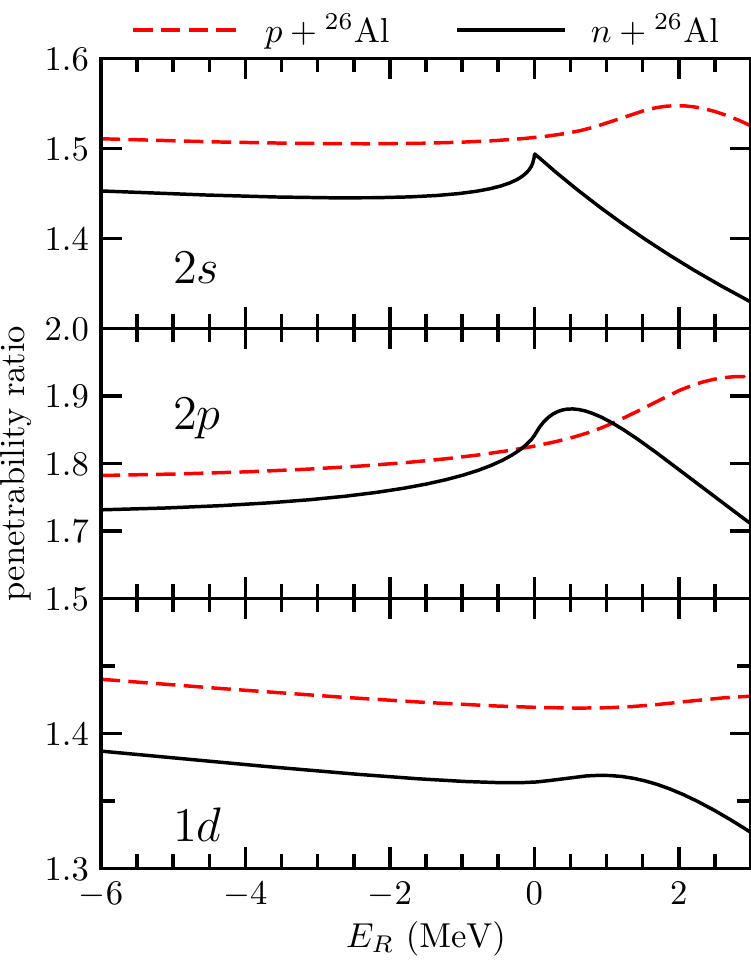}
\caption{The penetrability ratio as a function of the level energy
for $p+{}^{26}{\rm Al}$ and $n+{}^{26}{\rm Al}$ for $n\ell=2s$, $2p$, and $1d$.
See Subsec.~\ref{subsec:nuc_pen} for details.} \label{fig:pen_ratio_al26_N}
\end{figure}

By including the attractive tail of the nuclear potential in the
calculation of the Coulomb functions, the penetration factors are
increased compared to a Coulomb-only calculation.
The advantage of this approach is that single-particle ANCs and
widths calculated from potentials with a tail, such as the Woods-Saxon
potential used here, can be expressed using $R$-matrix formulas involving
single-particle reduced-width amplitudes, such as shown by
Eqs.~(\ref{eq:R_matrix_ANC}) and~(\ref{eq:R_matrix_width}).
Here, I consider the {\em penetrability ratio} defined to be the ratio
of the penetration factor calculated with the nuclear potential included to
that calculated without. This ratio can be defined for any energy
to be $|O(a)/\mathcal{O}(a)|^2$. When the energy is real and positive,
this becomes $\mathcal{P}(a)/P(a)$ since
$\mathcal{P}(a)=ka/[\mathcal{F}^2(a)+\mathcal{G}^2(a)]$ and
$P(a)=ka/[F^2(a)+G^2(a)]$ in this case.
For bound states, with $E$ real, the ratio is given by
$W^2(a)/\mathcal{W}^2(a)$.
This ratio is shown in Figs.~\ref{fig:pen_ratio_c12_N}
and~\ref{fig:pen_ratio_al26_N} for the cases of nucleon~+~${}^{12}{\rm C}$ and
nucleon~+~${}^{26}{\rm Al}$, for three $n\ell$ values and real energies.
For these calculations, the nuclear well depth has been adjusted to
place the single-particle level energy at the energy of the ratio
calculation, using the $R$-matrix boundary condition.
The ratios are seen to be moderately increased from unity,
reasonably independent of energy, and continuous across $E=0$.
The results for $n+{}^{12}{\rm C}$ are similar to those reported by
\textcite[Fig.~4]{Joh73} for $n+{}^{16}{\rm O}$ with a fixed well depth.

\subsection{Volume renormalization factor}
\label{subsec:volume_renorm}

\begin{figure}
\includegraphics[width=\columnwidth]{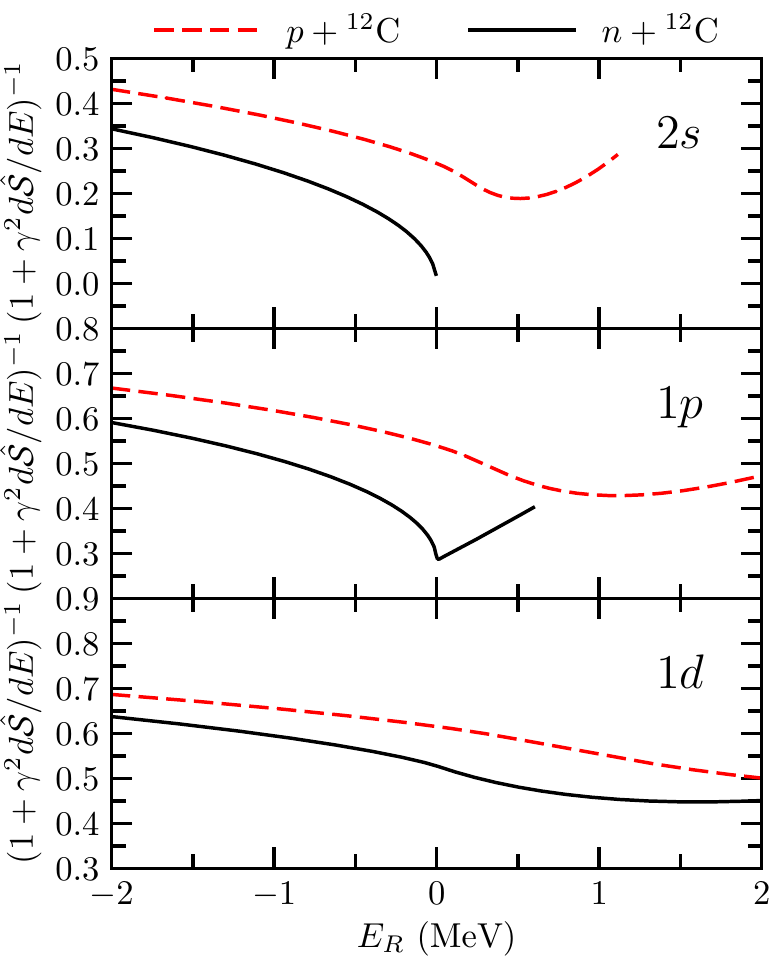}
\caption{The volume renormalization for $p+{}^{12}{\rm C}$ and
$n+{}^{12}{\rm C}$, for the single-channel case and the
single-particle reduced width.} \label{fig:factor_c12_N}
\end{figure}

\begin{figure}
\includegraphics[width=\columnwidth]{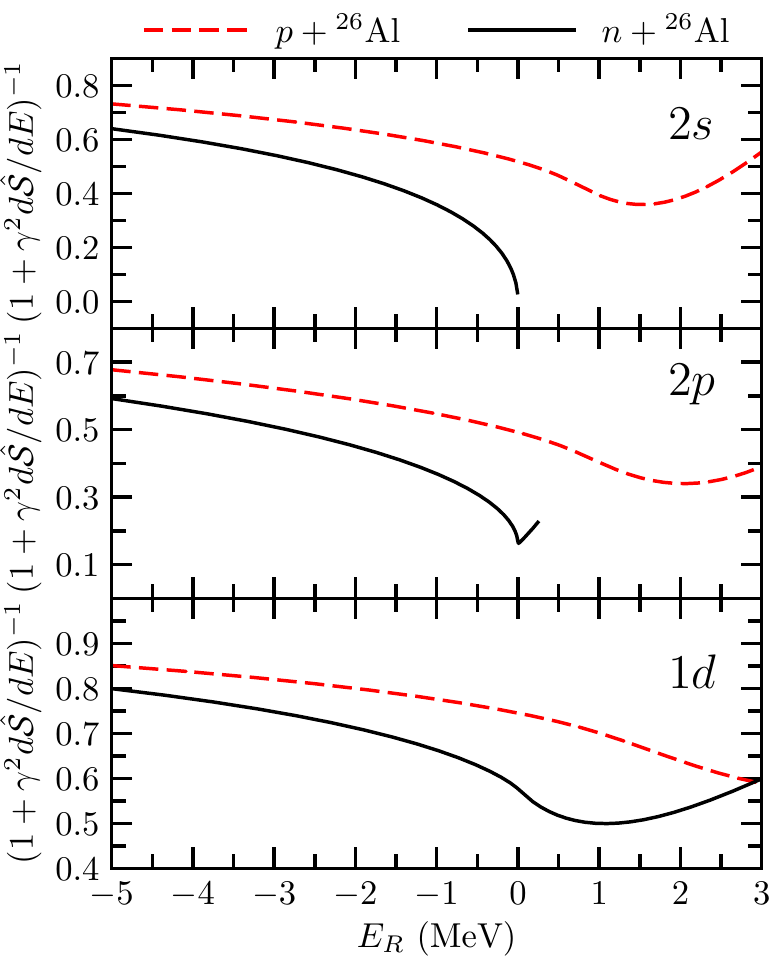}
\caption{The volume renormalization for $p+{}^{26}{\rm Al}$ and
$n+{}^{26}{\rm Al}$, for the single-channel case and the
single-particle reduced width.} \label{fig:factor_al26_N}
\end{figure}

In the single-channel case, the volume renormalization factor for
$R$-matrix states is given by
\begin{equation}
\left[1+\gamma^2\left(\frac{\partial\hat{\mathcal{S}}}{\partial E}\right)_a
  \right]^{-1},
\end{equation}
which appears in many places in this work, including the definitions of
single-particle ANCs, Eq.~(\ref{eq:R_matrix_ANC}) and
single-particle widths, Eq.~(\ref{eq:R_matrix_width}).
In the many-channel case, this quantity generalizes to
\begin{equation}
\left[1+\sum_{c}\gamma_{c}^2\left(\frac{\partial\hat{\mathcal{S}}_{c}}{\partial E}
  \right)_{a_{c}}\right]^{-1}.
\end{equation}
It likewise appears in several contexts below, including
Eqs.~(\ref{eq:ANC_reduced}), (\ref{eq:gamma_R_reduced}),
and~(\ref{eq:x_c_gamma_c}).
The volume renormalization factor for Gamow states, where $\hat{\mathcal{S}}$
is replaced by $\mathcal{L}$, is very similar, being identical in the
case of bound states and having a small complex component for
narrow resonances.
Examples of the volume renormalization factor for nucleon~+~${}^{12}{\rm C}$
and nucleon~+~${}^{26}{\rm Al}$ are shown in Figs.~\ref{fig:factor_c12_N}
and~\ref{fig:factor_al26_N}, for low partial waves.
These calculations are for the single-channel case and assuming the
single-particle reduced width.
The factor differs significantly from unity in all cases, with the
largest differences being near the nucleon separation threshold.
For $\ell=1$ neutrons, a cusp at the threshold is produced.

It should be noted that this factor appears in several other
contexts besides the ones discussed in this paper.
For example, it can be used to explain threshold anomaly~\cite{Lan70,Lan77}
observed in $(d,p)$ reactions on heavy nuclei.
It also arises in the explanation of why states with a significant
single-particle structure tend to be located near separation
thresholds~\cite{Bar64}. In both of these cases, the factor is describing
the excitation energy compression for physical resonant states
that is caused by a channel threshold.
A more detailed mathematical model for this compression is provided by
the transformation methods described in Appendix~\ref{app:transforms}
of this work. In this picture, the compression occurs when one
transforms from eigenstates satisfying constant logarithmic boundary conditions
to resonant boundary conditions.

\section{Overlaps, Spectroscopic Factors, and Reduced Width Amplitudes}
\label{sec:overlaps}

The radial overlap function $\mathcal{R}_c(r)$ is the projection of a many-body
nuclear wave function on to a particular channel
configuration~\cite{Ber65,Pin65,Tim03,Mic08,Tim10,Nol11,Tim11,Nol12,Muk19}
that satisfies an inhomogeneous Schr{\" o}digner-like equation.
For large $r$, $\mathcal{R}_c(r)$ satisfies a one-body radial
Schr{\" o}dinger equation with the intercluster Coulomb potential and
the cluster separation energy.
Here, channels denote a two-cluster configuration with quantum
numbers discussed in the first paragraph of Sec.~\ref{sec:single_particle}.
Channels will be labeled with the subscript $c$ when necessary.
The overlap function allows one to link together the spectroscopic
factor, reduced-width amplitude, and many-body theoretical calculations.

\subsection{Computational point of view}

If the many-body wave function $|\Psi\rangle$ is known, for example, by
finding the eigen solution to a given Hamiltonian,
then the overlap function may be calculated. One may then define
spectroscopic factors according to~\cite{Tho09}
\begin{equation} \label{eq:S_def}
\mathcal{S}_c = \int_0^\infty \mathcal{R}_c^2(r) \, r^2 \, dr ,
\end{equation}
where it assumed that the wave function is normalized to unity over all space,
i.e., $\langle\tilde{\Psi}|\Psi\rangle_0^\infty\equiv 1$.
Because of the normalization convention, the spectroscopic factor can only
be strictly defined for Gamow states.
In this work, I refer to multichannel states as Gamow states if they have
outgoing waves in all channels. This definition includes bound states.
It is also useful to define the spectroscopic amplitude whose square is
the spectroscopic factor
\begin{equation}
\mathcal{S}_c = \mathcal{A}_c^2 .
\end{equation}
Note that the spectroscopic factor and amplitude, as defined here, include an
isospin Clebsch-Gordan factor that is discussed below in
Sec.~\ref{sec:isospin}.
Time-reversal invariance allows $\mathcal{R}_c$, $\mathcal{S}_c$, and
$\mathcal{A}_c$ to be defined as real quantities when the energy of the
state is real.

The reduced-width amplitude is defined according
to~\cite[III.4, Eq.~(4.8a), p.~271]{Lan58}
\begin{equation} \label{eq:gamma_c}
\gamma_c = \left( \frac{\hbar^2 a_c}%
  {2\mu_c \langle\tilde{\Psi}|\Psi\rangle_0^a} \right)^{1/2} \mathcal{R}_c(a_c) ,
\end{equation}
where the factor $\langle\tilde{\Psi}|\Psi\rangle_0^a$ implements the
$R$-matrix convention that the wave function is normalized to unity {\em inside}
the channel radii.
When the energy is real, $\langle\tilde{\Psi}|\Psi\rangle_0^a$ is
real, positive, and less than one.
Because the normalization is not extended to infinity, reduced widths may
be defined for a much broader class of states, including unbound $R$-matrix
eigenstates.
The single-channel $R$-matrix resonance condition given in
Subsec.~\ref{subsec:r_matrix} generalizes naturally to
\begin{equation} \label{eq:R_boundary_cond}
\left[\frac{1}{\mathcal{R}_c}\frac{d(r\mathcal{R}_c)}{dr}\right]_{a_c} =
  \left[\hat{\mathcal{S}}_c\right]_{a_c} ,
\end{equation}
for all channels and at the level energy.
This represents the real part of the outgoing wave boundary condition.

It should be noted that spectroscopic factors are subject to some theoretical
ambiguity arising from from how the underlying nuclear interactions are
defined~\cite{Fri79,Muk10,Nol11}.
On the other hand, widths and ANCs, which are asymptotic quantities, are
free from such ambiguities. The reduced width amplitude, being nearly
asymptotic, is also essentially free from this issue.

\subsection{Phenomenological point of view}
\label{subsec:overlap_phenom}

In the phenomenological approach, neither the Hamiltonian nor the
wave function are assumed to be known. One instead works directly with
level energies, ANCs, and partial widths.
In addition, the radial overlap function may be approximated using the
replacement
\begin{equation} \label{eq:replace_overlap}
\mathcal{R}_c(r) \rightarrow \mathcal{A}_c \, \frac{u_c(r)}{r} ,
\end{equation}
where $u_c(r)/r$ is the single-particle radial wave function, as discussed in
Sec.~\ref{sec:single_particle}. Provided that $u_c(r)$ is normalized such
that $I_\infty=1$ via Eq.~(\ref{eq:I_infinity}), this replacement is
consistent with Eq.~(\ref{eq:S_def}). This approximation is commonly used
in transfer reaction calculations.
In bound channels, one then finds for the square of the ANC
\begin{equation} \label{eq:anc_spec}
C_c^2 = \mathcal{S}_c \, C^2_{c,\mathrm{sp}} ,
\end{equation}
where $C^2_{c,\mathrm{sp}}$ is the square of the single-particle ANC given by
Eq.~(\ref{eq:ANC_1}), (\ref{eq:ANC_2}), or (\ref{eq:R_matrix_ANC}).
For unbound channels, the partial width is given by
\begin{equation} \label{eq:width_spec}
\Gamma_c = \mathcal{S}_c \, \Gamma_{c,\mathrm{sp}} ,
\end{equation}
where $\Gamma_{c,\mathrm{sp}}$ is the single-particle partial width.
Strictly speaking, the single-particle partial width should be taken
as $\Gamma_{S1}$, as defined Eq.~(\ref{eq:Gamma_S1}), since $\mathcal{A}_c$
scales the asymptotic single-particle wave function.
However, as discussed in Subsec.~\ref{subsec:single_discuss}, all of the
width definitions are approximately equivalent if the single-particle
width is narrow.

In a phenomenological analysis, the number of channels is limited to
one or a small number. Those with a large spectroscopic factor,
small separation energy, and/or low orbital angular momentum are likely
to be important and should be included.
As discussed in Subsec.~\ref{subsec:coulomb}, it is
assumed that the Coulomb functions can be extended inward to the channel
radii, including the effects of the nuclear single-particle potential.
This approach has been utilized in several studies, including
Refs.~\cite{Tho68,Rob69,Wes71,Joh73,Koo74}.
While this procedure is a very reasonable approximation where the tail of
the nuclear potential is concerned, it does have some limitations.
Since the true overlap functions satisfy an inhomogeneous radial
equation (see, e.g., Ref.~\cite{Tim11}), there will be small violations
of the Wronskian relation satisfied by the modified Coulomb functions.
Also, \textcite[pp.~494-495]{Rob69} has noted that using channel radii just
outside the nuclear surface gives rise to some mild nonorthogonality
between the channels.
These effects could be removed by using larger channel radii, but
that would introduce additional breaking of isospin symmetry.
I agree with \textcite{Rob69} that channel radii just outside the
nuclear surface are the best choice for phenomenological analyses when
isospin symmetry is utilized.

In the phenomenological approach, the extension of $\mathcal{R}_c(r)$ for a
Gamow state beyond the channel radii is given by
\begin{subequations}
\begin{align}
\mathcal{R}_c(r) &= \mathcal{R}_c(a_c)
  \frac{\mathcal{O}_c(r)/r}{\mathcal{O}_c(a_c)/a_c} \\
     &= \gamma_c \left( \frac{2\mu_c \langle\tilde{\Psi}|\Psi\rangle_0^a}%
        {\hbar^2 a_c} \right)^{1/2}
        \frac{\mathcal{O}_c(r)/r}{\mathcal{O}_c(a_c)/a_c} ,
\end{align}
\end{subequations}
where Eq.~(\ref{eq:gamma_c}) has been utilized.
Equation~(\ref{eq:dlogude}) implies
\begin{equation}
  \int_{a_c}^\infty \frac{\mathcal{O}_c(r)}{\mathcal{O}_c(a_c)} \, dr =
  \frac{\hbar^2}{2\mu_c a_c} 
  \left(\frac{\partial\mathcal{L}_c}{\partial E}\right)_{a_c},
\end{equation}
which is a regularized value if the channel is unbound. One then has
\begin{equation}
  \int_{a_c}^\infty \mathcal{R}_c^2(r) \, r^2 \, dr =
  \gamma_c^2 \langle\tilde{\Psi}|\Psi\rangle_0^a
  \left(\frac{\partial\mathcal{L}_c}{\partial E}\right)_{a_c} .
\end{equation}
The normalization condition $\langle\tilde{\Psi}|\Psi\rangle_0^\infty\equiv 1$
may be expressed as
\begin{equation}
\langle\tilde{\Psi}|\Psi\rangle_0^a +
  \sum_c \int_{a_c}^\infty \mathcal{R}_c^2(r) \, r^2 \, dr = 1 ,
\end{equation}
which yields
\begin{equation}
\langle\tilde{\Psi}|\Psi\rangle_0^a = \left[1+\sum_c \gamma_c^2
  \left(\frac{\partial\mathcal{L}_c}{\partial E}\right)_{a_c} \right]^{-1}.
\end{equation}
This is a generalization of the well-known volume renormalization factor
in $R$-matrix
theory~\cite[IV.7, p. 280; Eqs. (A.29) and (A.30), p. 351]{Lan58}
that is discussed above in Subsec.~\ref{subsec:volume_renorm}.
Equations~(\ref{eq:reduced_width}) and~(\ref{eq:I_infinity}), with
$I_\infty=1$, yield
\begin{equation}
\frac{\hbar^2}{2\mu_c a_c} \frac{u_c^2(a_c)}{\gamma_{c,\mathrm{sp}}^2} =
\left[ 1+\gamma_{c,\mathrm{sp}}^2 \left(\frac{\partial\mathcal{L}_c}{\partial E}
\right)_{a_c} \right]^{-1} .
\end{equation}
The square of Eq.~(\ref{eq:gamma_c}), with the replacement
$\mathcal{R}_c^2(a_c)\rightarrow \mathcal{S}_c u_c^2(a_c)/a_c^2$, then provides
\begin{equation} \label{eq:spec_single}
\frac{\gamma_c^2}{1+\sum_{c'} \gamma_{c'}^2
  \left(\frac{\partial\mathcal{L}_{c'}}{\partial E}\right)_{a_{c'}}}
  = \mathcal{S}_c
  \frac{\gamma_{c,\mathrm{sp}}^2}{1+\gamma_{c,\mathrm{sp}}^2
  \left(\frac{\partial\mathcal{L}_c}{\partial E}\right)_{a_c}} .
\end{equation}
With these equations, it is straightforward to interoperate fully among
the single-particle wave functions and spectroscopic factors and among the
single-particle and actual ANCs, partial widths, and reduced widths.
For a state bound in channel $c$, Eqs.~(\ref{eq:R_matrix_ANC}),
(\ref{eq:anc_spec}), and~(\ref{eq:spec_single}) may be combined to yield
\begin{equation} \label{eq:ANC_reduced}
C_c^2 = \frac{2\mu_c a_c}{\hbar^2 \mathcal{W}_c^2(a_c)} \frac{\gamma_c^2}%
  {\rule{0pt}{1.0em} 1+\sum_{c'} \gamma_{c'}^2
  \left(\frac{\partial\hat{\mathcal{S}}_{c'}}{\partial E}\right)_{a_{c'}}} ,
\end{equation}
which is the general relation between the ANC and the reduced widths.
For a state that is unbound in channel $c$, one may likewise
combine Eqs.~(\ref{eq:Gamma_S1}), (\ref{eq:amplitude_reduced}),
and~(\ref{eq:spec_single}) to obtain
\begin{equation} \label{eq:gamma_S1_reduced}
\Gamma_c = 2  \left|\frac{k_ca_c}{\mathcal{O}_c^2(a_c)} \,
  \frac{\gamma_c^2}%
  {1+\sum_{c'}\gamma_{c'}^2\left(\frac{\partial\mathcal{L}_{c'}}{\partial E}
  \right)_{a_{c'}}} \right| ,
\end{equation}
where this is the partial width defined by the $S$-matrix pole residue.
The corresponding partial width for the $R$-matrix definition is
\begin{equation} \label{eq:gamma_R_reduced}
\Gamma_c = 2  \mathcal{P}_c(a_c) \,
  \frac{\gamma_c^2}%
  {1+\sum_{c'}\gamma_{c'}^2\left(\frac{\partial\hat{\mathcal{S}}_{c'}}{\partial E}
  \right)_{a_{c'}}} ,
\end{equation}
with all of the terms in this formula being real quantities.
In what follows, it is useful to unify the treatment of bound and unbound
channels by defining
\begin{equation} \label{eq:xc}
X_c = \left\{\begin{array}{l@{\hspace*{0.2in}}l}
  \dfrac{\hbar^2\mathcal{W}_c^2(a_c)C_c^2}{2\mu_ca_c} & \mbox{bound channel}
  \\[1em]
  \dfrac{\Gamma_c}{2\mathcal{P}_c(a_c)} & \mbox{unbound channel}
  \end{array} \right. ,
\end{equation}
where the $R$-matrix definition of the partial width is utilized.

It should be noted that low-energy nuclear physics experiments are insensitive
to short-range features of nuclear wave functions.
Consequently, neither the single-particle potential nor the
spectroscopic factor are well constrained from a phenomenological point of view.
However, ANCs, widths, and reduced widths, being asymptotic or nearly
asymptotic quantities, can be constrained by such
experiments. This observation implies that a certain combination of
spectroscopic factor and single-particle wave function, essentially
$\mathcal{S}_c^{1/2} u_c(r)$ at and beyond the nuclear surface, can be
well constrained.

\subsection{An alternative definition of the spectroscopic factor}

When working in an $R$-matrix framework with channel radii, it is
convenient to utilize an alternative definition of the spectroscopic
factor that only depends on the wave function inside the channel radii.
This property makes it very useful for studying isospin symmetry.
From a computational perspective, the alternative definition is
\begin{equation}
\mathbb{S}_c = \frac{\int_0^{a_c} \mathcal{R}_c^2(r) \, r^2 \, dr}%
  {\langle\tilde{\Psi}|\Psi\rangle_0^a} .
\end{equation}
In this work, I will refer to $\mathbb{S}_c$ as the {\em internal}
spectroscopic factor.
The phenomenological replacement of the radial overlap function,
analogous to Eq.~(\ref{eq:replace_overlap}), is
\begin{equation}
\frac{R_c(r)}{ \left(\langle\tilde{\Psi}|\Psi\rangle_0^a \right)^{1/2}}
  \rightarrow
  \mathbb{A}_c \frac{u_c(r)}{r \left(\int_0^{a_c} u_c^2(r) \, dr\right)^{1/2}} ,
\end{equation}
where $\mathbb{A}_c$ is the internal spectroscopic amplitude and
$\mathbb{S}_c=\mathbb{A}_c^2$.
The quantities $\mathbb{S}_c$ and $\mathbb{A}_c$ can again
be defined as real quantities when the energy is real.
Now, because the normalization does not extend over all space, the
internal spectroscopic factor and amplitude can be defined for a much
broader class of states without approximation, including $R$-matrix
eigenfunctions.
In this framework, the analog of Eq.~(\ref{eq:spec_single}) becomes
much simpler:
\begin{equation} \label{eq:spec_int_single}
\gamma_c^2 = \mathbb{S}_c\gamma_{c,\mathrm{sp}}^2 \quad\mbox{or}\quad
  \gamma_c = \mathbb{A}_c\gamma_{c,\mathrm{sp}} .  
\end{equation}

The difference between $\mathcal{S}_c$ and $\mathbb{S}_c$ is often small,
but this is not always the case, particularly when smaller channel radii
are utilized and/or when the state in question has low angular momentum
and lies near a channel threshold.
Also note that the difference disappears for single-particle states,
i.e., when $\mathcal{S}_c=\mathbb{S}_c=1$, indicating that differences
will be larger when the spectroscopic factors depart significantly from unity.
It is interesting to note that Eq.~(\ref{eq:spec_int_single}) is how
spectroscopic factors were originally defined~\cite{Fre60,Mac60}, but
this definition was largely supplanted by Eq.~(\ref{eq:S_def}).
The different definitions are alluded to in the work of
\textcite[pp.~489-490]{Rob69}.
These differences have led to some confusion in the
literature~\cite{Ili97,Bar98,Bar97,Moh97}.
It also appears that the denominators in Eq.~(\ref{eq:spec_single})
are sometimes dropped as an approximation.
It should be noted that the validity of such an approximation hinges in
part on {\em both} the single-particle and actual reduced widths being
sufficiently small.

Another consideration arises if spectroscopic factors from a shell model
calculation using harmonic oscillator basis states are utilized.
In this case, the energy eigenstates do not have the correct outgoing-wave
behavior beyond the channel radii. Instead, the magnitude of the
wave function falls off much more quickly with radius.
In this case, it is likely a better approximation to consider such
spectroscopic factors as {\em internal} spectroscopic factors 
$\mathbb{S}_c$ for the purpose of calculating ANCs or widths.

The distinction between $\mathcal{S}_c$ and $\mathbb{S}_c$ is closely
related to the distinction between observed and formal widths or reduced
widths; see \textcite[Sec.~5]{Des10} for definitions of these quantities.
In this work, all widths are defined to be observed widths and
all reduced widths to be formal reduced widths.
In addition, dimensionless reduced widths are not utilized in this work.
I find the proliferation of additional notation to be unnecessary and it also
creates additional opportunities for confusion.

\section{Isospin and mirror symmetry}
\label{sec:isospin}

Some examples of the use of isospin in the present context are provided
by Refs.~\cite{Tho68,Wer68,Rob69,Mon71,Hal90}.
If the nuclear state $B$ is a member of an isospin multiplet
with well-defined total isospin, its decay into clusters
$A$ and $a$, that are also assumed to have well-defined total isospins,
may be described using the isospin formalism.
It is assumed that $T_X$ are the total isospins of nuclei $X$, and
$T_{X3}$ are the corresponding isospin projections, where $X=B$, $A$, or $a$.
One then has for the spectroscopic amplitude~\cite[Eq.~(5.3.11), p.~193]{Tho09}
\begin{equation} \label{eq:spec_iso}
  \mathcal{A}_{c'} = \langle T_A T_{A3}, T_a T_{a3} | T_B T_{B3}\rangle
  \tilde{\mathcal{A}}_c .
\end{equation}
Alternatively, one can write
\begin{equation} \label{eq:spec_int_iso}
  \mathbb{A}_{c'} = \langle T_A T_{A3}, T_a T_{a3} | T_B T_{B3}\rangle
  \tilde{\mathbb{A}}_c
\end{equation}
or
\begin{equation} \label{eq:reduced_iso}
  \gamma_{c'} = \langle T_A T_{A3}, T_a T_{a3} | T_B T_{B3}\rangle
  \tilde{\gamma}_c .
\end{equation}
Note that in general a channel $c'$ can occur more than once in a particular
nucleus. For example, a $T=1$ $n+{}^3{\rm H}$ channel has both
$n+{}^3{\rm He}$ and $p+{}^3{\rm H}$ analogs in the ${}^4{\rm He}$ nucleus.
Mirror channels can only occur once in the respective nuclei.
In all three of the above cases, the spectroscopic amplitude or reduced width
written with the tilde symbol is common to the multiplet,
and the symbols on the left without the tilde vary across the multiplet,
depending upon the $T_{X3}$ values in the Clebsch-Gordan coefficient.
It is assumed that the other quantum numbers needed to define the channels
$c$ and $c'$ remain fixed across the multiplet.
It should also be noted that these definitions are in general not equivalent,
although the difference between the latter two is generally very small.
The latter two approaches can be made exactly equivalent if an
average single-particle reduced width is used for the
multiplet~\cite{Bar80,Bar92}.
In the case of an isospin mirror pair, the states have
opposite $T_{X3}$ components, resulting in spectroscopic factors
(for the first two definitions) or squared reduced width amplitudes
(for the third definition) that are equal.

Isospin symmetry is violated by the Coulomb interaction, which dominates
beyond the channel radii. It is further broken by energy displacements,
which also contribute to different radial dependences beyond the channel radii.
Consequently, one expects the first approach,
Eq.~(\ref{eq:spec_iso}) involving normalizations that extend to infinity,
to be less accurate than the second two, Eqs.~(\ref{eq:spec_int_iso})
and~(\ref{eq:reduced_iso}) involving normalizations inside the channel
radii~\cite{Rob69}.
Note also that the utilization of the more accurate approaches,
Eqs.~(\ref{eq:spec_int_iso}) or~(\ref{eq:reduced_iso}),
leads to isospin symmetry breaking in the
traditional spectroscopic factor defined by Eq.~(\ref{eq:spec_iso}).
As already discussed in Subsec.~\ref{subsec:overlap_phenom},
it is also important to utilize channel radii just outside the nuclear surface,
in order to avoid introducing additional isospin symmetry breaking.

If either of the first two approaches, defined by Eqs.~(\ref{eq:spec_iso})
and~(\ref{eq:spec_int_iso}) using spectroscopic amplitudes, are utilized
in conjunction with the well-depth procedure to determine the single-particle
wave functions, some additional dependence on the short-range behavior of
the single-particle potential is introduced.
For example, this procedure for determining the single-particle wave function
includes the Thomas-Ehrman shift~\cite{Tho52,Ehr51} in the energy.
However, if the level in question has a small spectroscopic amplitude
for the single-particle configuration, this energy shift is spurious.
It has also been found that nonlocal contributions to the single-particle
potential are important when the spectroscopic amplitude is small~\cite{Ber75}.
One approach to minimizing this issue that has been suggested is to
match the single-particle level energy by varying a surface potential
rather than the main (volume) Woods-Saxon potential~\cite{Win85}.
For the purposes of this work, the question can be bypassed
by adopting the third approach, defined by Eq.~(\ref{eq:reduced_iso})
using the reduced width amplitude. This procedure avoids making any
reference to properties of potentials or wave functions inside the channel
radii and will be utilized extensively in the remainder of this work.

\section{Single-level mirror symmetry}
\label{sec:single}

It is very common in practice to work with cases involving mirror symmetry
between isolated levels.
Here, I describe two approaches to this case and provide some examples.

\subsection{{\em R}-matrix approach}
\label{subsec:mirror_r_matrix}

If there is only a single important channel, the relationship
between ANCs and/or widths of the mirror states are particularly simple.
Using the $R$-matrix framework and Eq.~(\ref{eq:xc}),
the width or ANC of a level is related to the reduced width via
\begin{equation}
X^{-1}=\gamma^{-2}+\left(\frac{\partial\hat{\mathcal{S}}}{\partial E}\right)_a ,
\end{equation}
where the channel label has been dropped.
Assuming the value $\gamma^2$ is identical for the mirror levels in
question, as implied by Eq.~(\ref{eq:reduced_iso}), one then finds
\begin{equation} \label{eq:mirror_rmatrix}
X_1^{-1}-\left(\frac{\partial\hat{\mathcal{S}}_1}{\partial E}\right)_a =
X_2^{-1}-\left(\frac{\partial\hat{\mathcal{S}}_2}{\partial E}\right)_a ,
\end{equation}
where the subscript 1~or~2 indicates the particular member of the mirror pair,
keeping in mind that the states differ in both energy and charge.

The multichannel case is only slightly more complicated.
According to Eq.~(\ref{eq:xc}), one has
\begin{equation} \label{eq:x_c_gamma_c}
X_c = \frac{\gamma_c^2}%
  {1+\sum_{c'}\gamma_{c'}^2\left(\frac{\partial\hat{\mathcal{S}}_{c'}}{\partial E}
  \right)_{a_{c'}}} .
\end{equation}
This equation may be inverted to yield
\begin{equation} \label{eq:gamma_c_x_c}
\gamma_c^2 = \frac{X_c}{1-\sum_{c'}X_{c'}\left(
  \frac{\partial\hat{\mathcal{S}}_{c'}}{\partial E}
  \right)_{a_{c'}}} .
\end{equation}
Suppose the widths and/or ANCs $X_{1c}$ of a state in nucleus~1 are known.
The procedure is to determine corresponding widths and/or ANCs $X_{2c}$
of the mirror state~2 is as follows.
First, the $X_{1c}$ are converted into $\gamma_c$ using
Eq.~(\ref{eq:gamma_c_x_c}). Then the $\gamma_c$ are converted to $X_{2c}$
using Eq.~(\ref{eq:x_c_gamma_c}), implicitly assuming the reduced widths
$\gamma_c$ are equal for both states. The result of this procedure is
\begin{equation} \label{eq:R_mirror}
X_{2c} = \frac{X_{1c}}{\rule{0pt}{1.5em} 1+\sum_{c'}X_{1c'}\left[ 
  \left( \frac{\partial\hat{\mathcal{S}}_{2c'}}{\partial E}  \right)
 -\left( \frac{\partial\hat{\mathcal{S}}_{1c'}}{\partial E}  \right)
  \right]_{a_{c'}} } .
\end{equation}

\subsection{Result of Timofeyuk and collaborators}
\label{subsec:timo}

Another formula relating ANCs and/or widths of mirror states has been
put forward by Timofeyuk and
collaborators~\cite{Tim03,Tim05a,Tim05b,Tim07,Muk19}.
It only considers a single channel and in the present notation reads
\begin{equation} \label{eq:r_timo}
\frac{X_1 / |\mathcal{O}_1(a)|^2}{X_2 / |\mathcal{O}_2(a)|^2} =
\left| \frac{\exp(i\sigma_1) \, F_1(\tilde{a}) / k_1 }%
            {\rule{0pt}{1.0em}
             \exp(i\sigma_2) \, F_2(\tilde{a}) / k_2 } \right|^2.
\end{equation}
Here, the quantity $\tilde{a}$ is a channel radius, but it need not be
the same as $a$. The left side of this equation is essentially a ratio
of ANCs and/or widths and the right side is a prediction.
This formula is nontrivially different from Eq.~(\ref{eq:mirror_rmatrix}),
that does not involve the regular Coulomb function or predict the
relationship to be a ratio.
As discussed in Refs.~\cite{Tim03,Tim07}, the derivation of this formula
depends on certain assumptions about the wave functions and
matrix elements of the Coulomb interaction in the nuclear interior.

Some insight into this equation can be deduced in the $R$-matrix framework by
making some assumptions regarding Eqs.~(\ref{eq:asymp_integral})
and~(\ref{eq:r_int_asymp}), for $r\le\tilde{a}$:
\begin{subequations} \label{eq:internal_assume}
\begin{align}
&\frac{F_1(r)}{\rule{0pt}{1.0em} F_1(\tilde{a})} =
 \frac{F_2(r)}{\rule{0pt}{1.0em} F_2(\tilde{a})} , \label{eq:timo_assump1} \\
&\begin{aligned}[c]
  [V(r)+&V_C(r)- V_{pC}(r)]_1 = \\ & [V(r)+V_C(r)-V_{pC}(r)]_2 ,
\end{aligned} \label{eq:timo_assump2} \\
&u_1(r) = u_2(r) , \mbox{~and} \label{eq:mirror_rad_wf} \\
&b=a=\tilde{a} .
\end{align}
\end{subequations}
Note that $u_1(r)=u_2(r)$ embodies the mirror-symmetry assumption and
implies the single-particle reduced widths are equal.
The assumption that $b=a$ implies that the channel radius is large
enough such that at the channel radius, Coulomb interactions are negligible
and the unmodified Coulomb functions can be utilized.
In Eq.~(\ref{eq:r_int_asymp}), I further assume the unbound state is well
below the Coulomb and angular momentum barriers such that
$G(a)\gg F(a)$ and the left-hand side of the equation may be replaced
by $u(a)[P(a)/(ka)]^{1/2}$.
Then, Eq.~(\ref{eq:asymp_integral}) (for a bound state) and
Eq.~(\ref{eq:r_int_asymp}) (for an unbound state) both lead to
\begin{equation}
\begin{split}
\gamma^2_{\mathrm{sp}} = & \frac{2\mu|O(a)|^2}{\hbar^2 a} \\
  & \times\left| \frac{\exp(i\sigma)}{k}\int_0^a F(V+V_C-V_{pC})\, u \, dr
  \right|^2 .
\end{split}
\end{equation}

Considering the assumptions described by Eq.~(\ref{eq:internal_assume})
and assuming equal internal spectroscopic factors described by
Eq.~(\ref{eq:spec_int_single}), the quantity
\begin{equation} \label{eq:timo_factor}
\left| O(a) F(a) \frac{\exp(i\sigma)}{k} \right|^2
\end{equation}
should be equal for both members of the mirror pair.
If this is true, then Eq.~(\ref{eq:r_timo}) reduces to
$X_1=X_2$, which is equivalent to Eq.~(\ref{eq:mirror_rmatrix}), {\em if} the
volume renormalization factors $\partial\hat{S}/\partial E$ are neglected.
In fact, \textcite{*[{}] [{. Note that the ${}^\prime$ in Eq.~(20) of this
paper indicates differentiation with respect energy, while the ${}^\prime$
in the following equation that defines $S_l$ indicates differentiation
with respect to $\kappa a$.}] Tim05b} point out that
Eq.~(\ref{eq:r_timo}) should
be modified by the volume renormalization factor for the case of an unbound
state. However, they mention no such correction for bound states,
although it is clear that it should be included in this case as well.

A mathematical explanation of why the quantity given by
Eq.~(\ref{eq:timo_factor}) is approximately equal for both members is provided
by the Wentzel-Kramers-Brillouin (WKB) approximation~\cite{Sch68}.
For bound states, or unbound states
below the Coulomb and/or angular momentum barriers, the solutions to
Eq.~(\ref{eq:schrodinger}) for radii beyond the range of the nuclear
potential depend exponentially on the radius.
Following Ref.~\cite[Eqs.~(34.4), (34.5), and~(34.8), pp.~270-271]{Sch68},
\begin{subequations}
\begin{align}
u_{\pm} &\propto \kappa^{-1/2}\exp(\pm{\textstyle \int} \kappa\,dr )
  \quad\quad\mbox{with} \\
\kappa &=\left\{ \frac{2\mu}{\hbar^2}\left[-E+\frac{\hbar^2}{2\mu}
  \frac{\ell(\ell+1)}{r^2}+V_c(r) \right]\right\}^{1/2} , \label{eq:kappa}
\end{align}
\end{subequations}
where $\kappa$ is real. One also finds
\begin{equation}
\frac{1}{u_\pm}\frac{du_\pm}{dr} = -\frac{1}{2\kappa}\frac{d\kappa}{dr}\pm\kappa
\end{equation}
for the logarithmic derivative. For the energy regime under consideration,
the regular Coulomb function $F$ is identified as the exponentially-increasing
solution and the outgoing Coulomb function $O$ as the exponentially-decreasing
solution. One thus has
\begin{subequations}
\begin{align}
\frac{1}{F}\frac{dF}{dr} & \approx
\left(\frac{1}{u_+}\frac{du_+}{dr}\right)_{\rm WKB} =
  -\frac{1}{2\kappa}\frac{d\kappa}{dr}+\kappa \quad\mbox{and} \\
\frac{1}{O}\frac{dO}{dr} & \approx
\left(\frac{1}{u_-}\frac{du_-}{dr}\right)_{\rm WKB} =
  -\frac{1}{2\kappa}\frac{d\kappa}{dr}-\kappa  .
\end{align}
\end{subequations}
I note in passing that this result for the WKB shift function
\begin{equation}
\hat{S}_{\rm WKB} = \left(\frac{r}{u_-}\frac{du_-}{dr}\right)_{\rm WKB} =
  -\frac{r}{2\kappa}\frac{d\kappa}{dr}-\kappa r
\end{equation}
agrees with that given in \textcite[Eq.~(A.18), p.~350]{Lan58},
apart from their Langer modification.
As discussed in Ref.~\cite{Lan58}, this expression may
also be used to derive a WKB approximation for the energy derivative
of the shift function. The Wronskian relation, Eq.~(\ref{eq:wronskian}),
may be written as
\begin{equation}
\frac{1}{F}\frac{dF}{dr}-\frac{1}{O}\frac{dO}{dr} = \frac{k}{FO\exp(i\sigma)}.
\end{equation}
Thus, in the WKB approximation,
\begin{equation} \label{eq:wkb_wronskian}
\left| O(a) F(a) \frac{\exp(i\sigma)}{k} \right|_{\rm WKB}= \frac{1}{2\kappa}.
\end{equation}
One is now in a position to understand why this quantity will be approximately
the same for mirror states. Considering Eq.~(\ref{eq:schrodinger})
and the approximation given by Eq.~(\ref{eq:mirror_rad_wf}) at $r=a$, one finds
\begin{equation}
E_2-E_1 \approx V_{C2}(a) - V_{C1}(a)
\end{equation}
for the Coulomb energy difference of the single-particle wave functions.
Then considering Eq.~(\ref{eq:kappa}), one has $\kappa_1(a)\approx\kappa_2(a)$,
and one does indeed find Eq.~(\ref{eq:wkb_wronskian}) to be the same
for both states of the mirror pair.
It is important to note that the Coulomb energy difference plays a key
role in this approximate equivalence and that of
Eqs.~(\ref{eq:mirror_rmatrix}) and~(\ref{eq:r_timo}).
A somewhat similar analysis of the justification for Eq.~(\ref{eq:r_timo}) has
been given in Ref.~\cite{Tim07}.

\subsection{Discussion}

Four methods for implementing mirror symmetry have been introduced.
Three of the methods are based on Eqs.~(\ref{eq:spec_iso}),
(\ref{eq:spec_int_iso}), and~(\ref{eq:reduced_iso}); the fourth
is described in the previous subsection.
The $R$-matrix approach based on Eq.~(\ref{eq:reduced_iso}) is
described in detail in Subsec.~\ref{subsec:mirror_r_matrix}.
Of all of the approaches, this one most strongly adopts the spirit
of the phenomenological $R$~matrix, as no assumptions about potentials
or wave functions inside the channel radii are necessary.
Approaches based upon Eq.~(\ref{eq:spec_iso}) are expected to be
somewhat less accurate than the others, because this spectroscopic
amplitude is normalized over all space, which unnecessarily includes isospin
violation due to the Coulomb force beyond the channel radii.
Using the internal spectroscopic amplitude, as defined by
Eq.~(\ref{eq:spec_int_iso}), does not suffer from this shortcoming.
It can also supply some internal mirror symmetry
breaking due to mirror symmetry breaking in the single-particle
reduced-width amplitudes.
The $R$-matrix approaches, Eq.~(\ref{eq:spec_int_iso})
or~(\ref{eq:reduced_iso}), also have the advantage of allowing multichannel
effects to be included.
It is not clear from this discussion which
approach, Eq.~(\ref{eq:spec_int_iso}) or~(\ref{eq:reduced_iso}), is
preferable. It may be possible to address this question in particular
cases if accurate many-body calculations are available.
I have some preference for Eq.~(\ref{eq:reduced_iso}), due to its
conceptual simplicity.

The approach of Timofeyuk described in Subsec.~\ref{subsec:timo}
leads to results that are similar to the other methods in most cases.
However, this formula does not take into account the volume renormalization
factors, which can lead to a significant error if these factors
differ significantly from unity.
This consideration is particularly relevant for the first excited
$1/2^+$ states of  ${}^{13}{\rm C}$ and ${}^{13}{\rm N}$ discussed below.

\subsection{Examples}

\subsubsection{\texorpdfstring{$\ell=0$}{l=0} mirror states in
  \texorpdfstring{${}^{13}{\rm C}$}{13C} and
   \texorpdfstring{${}^{13}{\rm N}$}{13N} }

The first excited $1/2^+$ states of ${}^{13}{\rm C}$ and ${}^{13}{\rm N}$
have long been a testing ground for mirror symmetry~\cite{Tho52,Bar80,Tim05b}.
The level is single~particle in nature and is bound in ${}^{13}{\rm C}$
but unbound in ${}^{13}{\rm N}$.
The neutron ANC in ${}^{13}{\rm C}$ has been measured by two independent
groups using the ${}^{12}{\rm C}(d,p){}^{13}{\rm C}$ transfer reaction in
similar kinematics. \textcite{Liu01} measured
$C^2_n=3.39\pm 0.59$~(stat~+~sys)~fm${}^{-1}$, while \textcite{Ima01} reported
$C^2_n=3.65\pm 0.34$~(stat)~$\pm 0.35$~(sys)~fm${}^{-1}$.
In neither experiment is it clear if the systematic uncertainty includes
the theoretical uncertainty from the transfer reaction analysis;
I adopt $C^2_n=3.52\pm0.50$~fm${}^{-1}$.
The proton width in ${}^{13}{\rm C}$ is taken from the elastic scattering
data and $R$-matrix analysis of \textcite{Mey76}.
This work does not quote uncertainties; I adopt $\Gamma_p=33.8\pm 2.0$~keV,
which is also consistent with the analysis of Ref.~\cite{Bar80}.

\begin{figure}
\includegraphics[width=\columnwidth]{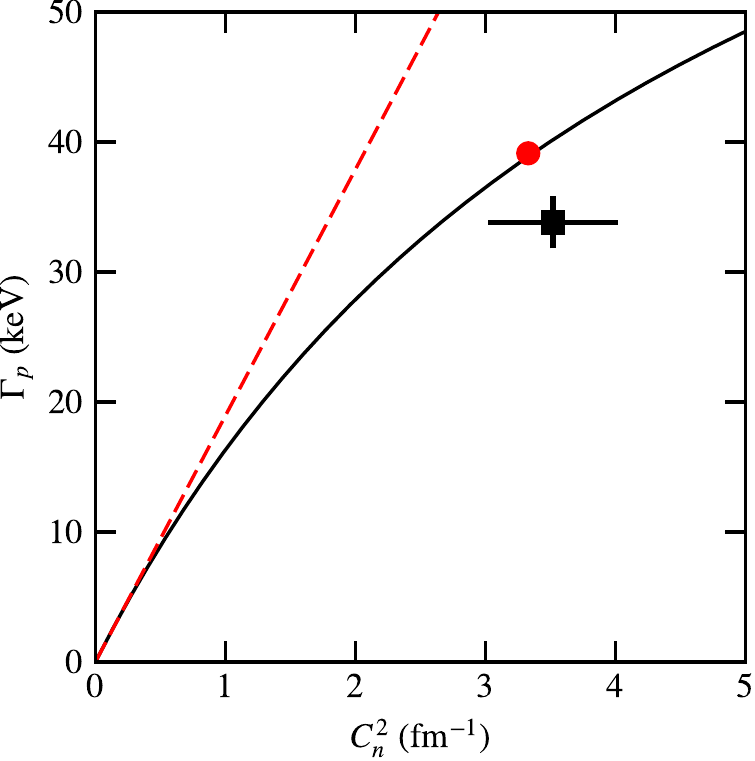}
\caption{The relationship between $\Gamma_p$ and $C^2_n$ for the
first excited states of ${}^{13}{\rm C}$ and ${}^{13}{\rm N}$.
The solid and dashed curves show the results of Eqs.~(\ref{eq:mirror_rmatrix})
and~(\ref{eq:r_timo}), respectively. The filled circle shows the
results of the single-particle calculation. The filled square indicates
the experimental results. } \label{fig:analyze-c12N}
\end{figure}

\begin{figure}
\includegraphics[width=\columnwidth]{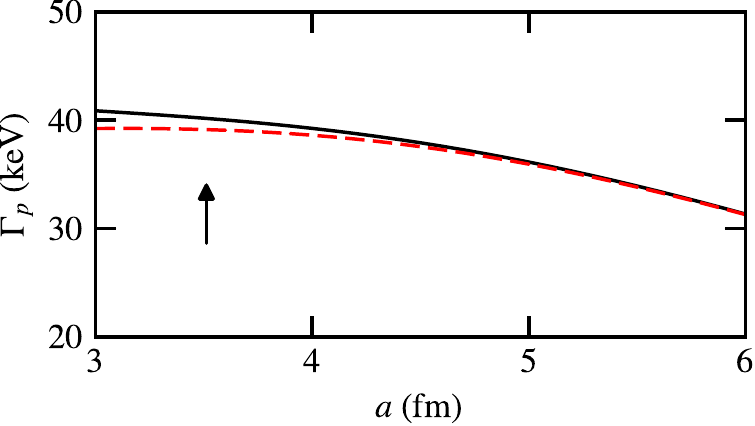}
\caption{The relationship between $\Gamma_p$ and the
channel radius predicted by Eq.~(\ref{eq:mirror_rmatrix})
for a fixed value of $C_n^2$. The solid and dashed curves show the results
including of including and ignoring the tail of the\
nuclear potential, respectively.
The arrow indicates the nominal channel radius of 3.51~fm used for
the calculations shown in Fig.~\ref{fig:analyze-c12N} and discussed
in the text.} \label{fig:analyze-c12N-chan-rad}
\end{figure}

First, calculations were performed using two-body potential and assuming
spectroscopic factors of unity. The potential depth was adjusted separately
for each state to reproduce the known separation energy. With the standard
potential parameters given in Subsec.~\ref{subsec:practical},
the depths for each state only difference by a few percent.
Likewise, the single-particle reduced widths for each state
only differed by a few percent. The resulting ANC and proton width are shown
as the filled circle in Fig.~\ref{fig:analyze-c12N}.
Then the potential was fixed at a depth taken to be the average
two results found for each state.
The relationship predicted for $\Gamma_p$ versus $C^2_n$ using
Eq.~(\ref{eq:mirror_rmatrix}), taking the common $\gamma^2$ to be a
varying parameter, is shown as the solid curve in Fig.~\ref{fig:analyze-c12N}.
The curve shows significant curvature, due to volume renormalization factors
that depend upon $\partial\hat{\mathcal{S}}/\partial E$.
The experimental results are shown as the
filled square with error bars, which are in fair agreement with
the solid curve.

The sensitivity of Eq.~(\ref{eq:mirror_rmatrix}) to the channel radius is
shown as the solid curve in Fig.~\ref{fig:analyze-c12N-chan-rad},
for $C_n^2=3.52$~fm${}^{-1}$.
This sensitivity is seen to be rather modest.
Equation~(\ref{eq:mirror_rmatrix}) is insensitive to the tail
of the nuclear potential: $\pm 10$\% changes in $r_n$ only change
the solid curve by 1\%.
Such changes do of course modify the single-particle ANC and width
more significantly.
The effect of ignoring the tail of nuclear potential completely is
shown as the dashed curve. This insensitivity indicates that the
tail of the nuclear potential could be safely ignored for this calculation.

The prediction of Eq.~(\ref{eq:r_timo}) assuming the same channel
radius is shown by the dashed curve
in Fig.~\ref{fig:analyze-c12N}. For larger values of $C_n^2$ and $\Gamma_p$,
it is seen to diverge significantly from the solid curve given by
Eq.~(\ref{eq:mirror_rmatrix}), the single-particle values, and
the experimental measurements.
The disagreement between Eq.~(\ref{eq:r_timo}) and the single-particle model,
microscopic models, and experiment has been noted
previously~\cite{Tim03,Tim05b,Muk19}.
If volume renormalization factors in
Eq.~(\ref{eq:mirror_rmatrix}) are neglected, the result from that
equation becomes very close to that Eq.~(\ref{eq:r_timo}).
I thus conclude that the disagreement between Eq.~(\ref{eq:r_timo})
and other approaches and experiment is due to the lack of volume
renormalization factors in Eq.~(\ref{eq:r_timo}), a deficiency that
has already been noted.

\subsubsection{\texorpdfstring{$\ell=0$}{l=0} mirror states in
  \texorpdfstring{${}^{17}{\rm O}$}{17O} and
   \texorpdfstring{${}^{17}{\rm F}$}{17F} }

The situation with the first excited $1/2^+$ states of ${}^{17}{\rm O}$ and
${}^{17}{\rm F}$ is quite similar to the previous example.
The states are single particle in nature, but in this case both states
are bound.
The neutron ANC in ${}^{17}{\rm O}$ has been determined by the analysis of
${}^{16}{\rm O}(d,p)$ data by \textcite{Guo07} to be
$C_n^2=8.4\pm 1.3$~fm${}^{-1}$.
The proton ANC in ${}^{17}{\rm F}$ has been reviewed by \textcite{Art00},
where their own and previous proton transfer experiments were
analyzed to yield $C_p^2=6220\pm 780$~fm${}^{-1}$.
The proton ANC was also determined using ${}^{16}{\rm O}({}^3{\rm He},d)$
by \textcite{Gag99} to be $C_p^2=6490\pm 680$~fm${}^{-1}$.
I adopt $C_p^2=6380\pm 510$~fm${}^{-1}$, which is also in the range
required to correctly describe low-energy ${}^{16}{\rm O}(p,\gamma)$
cross section measurements to the first excited state of
${}^{17}{\rm F}$~\cite{Bru96,Guo07,Ili08}.

\begin{figure}
\includegraphics[width=\columnwidth]{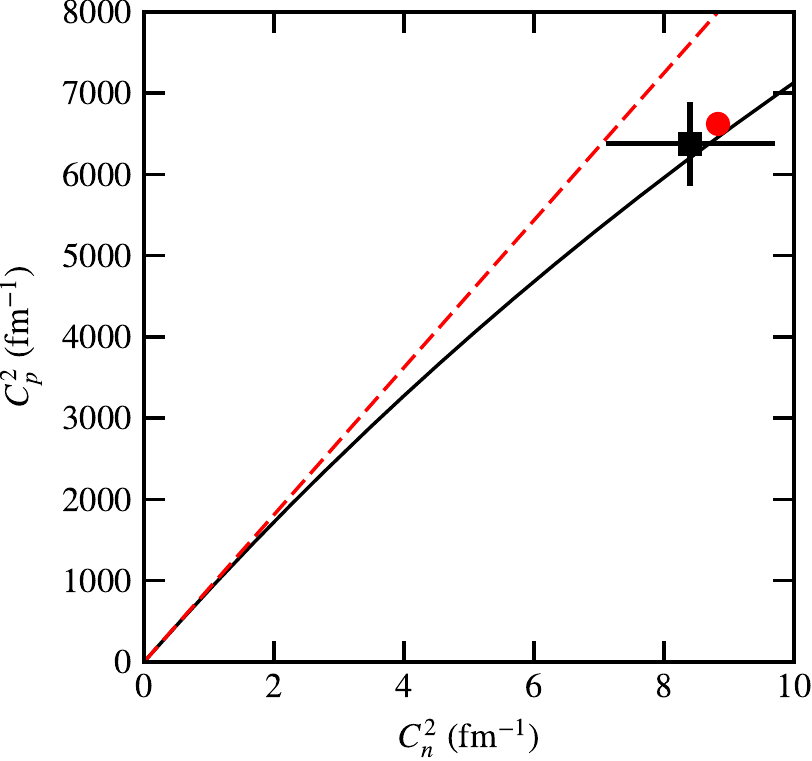}
\caption{The relationship between $C_p^2$ and $C^2_n$ for the
first excited states of ${}^{17}{\rm O}$ and ${}^{17}{\rm F}$.
The solid and dashed curves show the results of Eqs.~(\ref{eq:mirror_rmatrix})
and~(\ref{eq:r_timo}), respectively. The filled circle shows the
results of the single-particle calculation. The filled square indicates
the experimental results. } \label{fig:analyze-o16N}
\end{figure}

Calculations were performed in the same manner as in the previous example
and are shown in Fig.~\ref{fig:analyze-o16N}.
In this case, the prediction of Eq.~(\ref{eq:mirror_rmatrix}) does not
deviate so much from that of Eq.~(\ref{eq:r_timo}). This finding
results because this case is more tightly bound and has a higher charge,
leading to a smaller effect from the volume renormalization factors.
Both calculations are in reasonable agreement with the experimental results.

For this case, it has been noted by Refs.~\cite{Tim05a,Tim06,Tit11}
that Eq.~(\ref{eq:r_timo}) is not in good agreement with calculations using
the single-particle model or other more sophisticated models.
Possible explanations, such as core excitations, are discussed in these works.
However, the inclusion of the volume renormalization factors brings
Eq.~(\ref{eq:r_timo}) into much better agreement with the other models.
This appears to be the primary reason for the discrepancy.

\subsubsection{\texorpdfstring{$\ell=0$}{l=0} mirror states in
  \texorpdfstring{${}^{27}{\rm Al}$}{27Al} and
  \texorpdfstring{${}^{27}{\rm Si}$}{27Si} }

\begin{figure}
\includegraphics[width=\columnwidth]{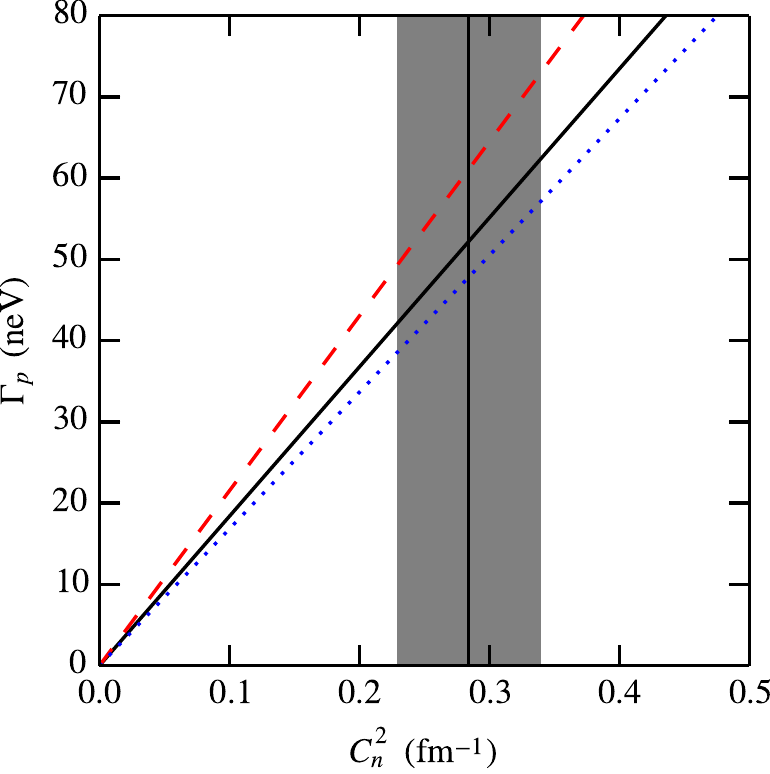}
\caption{The relationship between  $C^2_n$ and $\Gamma_p$ for the
7807- and 7590-keV $9/2^+$ states
of ${}^{27}{\rm Al}$ and ${}^{27}{\rm Si}$.
The solid, red-dashed, and blue-dotted curves show the results of
Eq.~(\ref{eq:mirror_rmatrix}), Eq.~(\ref{eq:spec_int_single}),
Eq.~(\ref{eq:anc_spec}), and Eq.~(\ref{eq:width_spec}), respectively.
The vertical line and gray error band show the adopted experimental
$C^2_n$ value. } \label{fig:analyze-al26N}
\end{figure}

The $9/2^+$ mirror pair located at $E_x=7807$~keV in $^{27}{\rm Al}$ and
7590~keV in ${}^{27}{\rm Si}$ couple to a nucleon and ${}^{26}{\rm Al}$
with $\ell=0$, but with a small spectroscopic factor of about 0.01.
This situation thus provides an example in a regime where the levels
in question are not close to being single-particle states.
Two independent measurements of the ${}^{26}{\rm Al}(d,p)$ reaction
are described in Refs.~\cite{Pai15,Mar15,Mar16,Lot20}.
Using the reported spectroscopic factors and neutron binding potentials,
the $\ell=0$ ANC value from Ref.~\cite{Pai15} is
$C^2_n=0.301\pm 0.062$~fm${}^{-1}$, where the error includes the experimental
uncertainty and a 15\% uncertainty from the transfer reaction analysis.
Similarly, Refs.~\cite{Mar15,Mar16} yield $C^2_n=0.259\pm 0.053$~fm${}^{-1}$,
where the error includes the experimental uncertainty only.
Since both experiments were performed with similar kinematics and utilized
nearly identical transfer reaction analyses, a common systematic error
of 15\% from the transfer reaction analysis is assumed for both
experiments, leading to an adopted value of $C^2_n=0.284\pm 0.054$~fm${}^{-1}$.
The mirror level is unbound in ${}^{27}{\rm Si}$, appearing as a resonance
at $E_R=126.8\pm 0.9$~keV.
Since $\Gamma_p \ll\Gamma_\gamma$ for this resonance, its strength is governed
by $\Gamma_p$ which can be estimated from $C^2_n$ of the mirror state.
This procedure has been carried out in Ref.~\cite{Pai15,Mar15,Mar16},
where it is found that this resonance dominates the
${}^{26}{\rm Al}(p,\gamma)$ reaction rate for temperatures
relevant to asymptotic giant branch and Wolf-Rayet stars.
The adopted resonance energy is determined from the
excitation energy measured by \textcite{Lot11} and the proton separation
energy from Ref.~\cite{Wan17}. Because the resonance is far below the
Coulomb barrier, the calculated $\Gamma_p$ is quite sensitive to the
energy: The 0.9-keV uncertainty contributes a 13\% uncertainty to
the $\Gamma_p$ deduced using mirror symmetry. Note also that this uncertainty
in the resonance energy contributes a further {\em correlated} uncertainty
in the thermonuclear reaction rate.

Although these levels can couple to $\ell=2$ nucleons, the contributions
of these channels negligibly effect the volume renormalization factors
and are neglected. Because of the small $\ell=0$ spectroscopic factor,
the volume renormalization factor is likewise small, leading to a linear
proportionality between $\Gamma_p$ and $C^2_n$ in all approaches.
Note, however,  that this factor cannot be neglected when calculating
single-particle ANCs or widths.
The predicted relation assuming equal reduced-width
amplitudes, Eq.~(\ref{eq:mirror_rmatrix}), is shown by the solid curve
in Fig.~\ref{fig:analyze-al26N}. This case has significant mirror
symmetry breaking in the single-particle reduced width amplitudes:
$\gamma^2_{n,\mathrm{sp}}=2.10$~MeV and $\gamma^2_{p,\mathrm{sp}}=2.46$~MeV,
a 17\% difference. Assuming a constant {\em internal} spectroscopic
factor and using Eq.~(\ref{eq:spec_int_single}) thus lead to a somewhat
different prediction, shown by the red-dashed line.
The approach of Timofeyuk and collaborators, using the same channel radius as
in the other approaches, is not shown but is very close to the red-dashed line.
Some previous analyses have assumed that the traditional spectroscopic factor
is the same for both states, and related $\Gamma_p$ and $C^2_n$ using
Eqs.~(\ref{eq:anc_spec}) and~(\ref{eq:width_spec}). As discussed in 
Sec.~\ref{sec:isospin}, this approach is expected to be somewhat less
accurate than the other two shown in Fig.~\ref{fig:analyze-al26N}.
This prediction is shown by the blue-dotted curve, where it is seen to lie
somewhat below the other two. The adopted experimental value for $C^2_n$
is shown by the vertical line and gray error band.
All of the approaches are in reasonable overall agreement and the
interpretation of the experimental data is not seriously limited by the
choice of model. Considering Eqs.~(\ref{eq:mirror_rmatrix})
and~(\ref{eq:spec_int_single}), the solid-black and red-dashed curves,
a value of $\Gamma_p=57\pm 15$~neV is extracted, in agreement with
previous determinations~\cite{Pai15,Mar15,Mar16}.
Note also that any deviation of a prediction from the blue-dotted curve
can be interpreted as a renormalization of the traditional spectroscopic
factor between the mirror states.

\section{Multilevel mirror symmetry}
\label{sec:multi}

It is important to note that all of the definitions of bound
or unbound resonant energy levels discussed up to this point violate isospin.
This occurs because the resonance condition, which is always some version
on an outgoing-wave boundary condition, depends upon the energy and charges
in the external region.
Since isospin rotations generally involve both changes in charge and
energy shifts, this situation is both necessary and expected.
In the $R$-matrix case, the resonance condition is given by
Eq.~(\ref{eq:R_boundary_cond}), which requires the logarithmic
derivative match the shift function at the channel radius.
The symmetry breaking coming from the boundary condition has
important effects if one considers isospin transformations on a set
of levels with the same spin and parity.
An $R$-matrix approach to calculating the effects arising from this
symmetry breaking is given below for the case of mirror symmetry,
along with an example application to $2^+$ states in
${}^{18}{\rm O}$ and ${}^{18}{\rm Ne}$.

\subsection{General phenomenological approach}
\label{subsec:gen_phen_mirror}

For examples of the use of isospin in multilevel phenomenological $R$-matrix
analyses, one may see Refs.~\cite{Wer68,Hal90} for light nuclei and
Ref.~\cite{Rob69} for heavier nuclei.
It is very useful to work in a basis that satisfies boundary conditions
that are independent of energy and isospin. 
The energy-independent boundary conditions of traditional $R$-matrix
theory~\cite{Wig47,Lan58} provide an ideal basis for this purpose.
For these states, the logarithmic derivatives at the channel radius are
equal to the constants $B_c$, rather than the energy- and charge-dependent
shift function of the resonance boundary conditions given by
Eq.~(\ref{eq:R_boundary_cond}).
Note also that any tail of the nuclear potential must be kept independent
of energy (i.e., fixed) in an $R$-matrix calculation.

I will consider the transformation from a set of resonance levels
of particular $J^\pi$ in nucleus~1 to a mirror nucleus~2.
A set of states $\{E_i(1),\gamma_{ic}(1)\}$ satisfying resonance boundary
conditions in nucleus~1 may be transformed into a set
$\{\hat{E}_\lambda(1),\hat{\gamma}_{\lambda c}(1), B_c\}$
satisfying constant boundary conditions as described in
Appendix~\ref{app:transforms}. The number of levels is preserved by
the transformation. I next suppose that the difference between the
Hamiltonians of the nuclei is $\Delta H=H_2-H_1$. Using the internal basis
states $|\lambda\rangle$ solving $H_1$ with boundary conditions $B_c$,
the level matrix~\cite[IX.1, Eq.~(1.11), p.~294]{Lan58}
for nucleus~2 may be written as
\begin{equation}
\begin{split}
[\bm{A}^{-1}]_{\lambda\mu} = & (\hat{E}_\lambda(1)-E)\delta_{\lambda\mu}
  +\langle\lambda|\Delta H|\mu\rangle \\ &
  -\sum_c \hat{\gamma}_{\lambda c}(1)\hat{\gamma}_{\mu c}(1)
  (\hat{\mathcal{S}}_c+i\mathcal{P}_c-B_c) ,
\end{split}
\end{equation}
where shift and penetration factors are evaluated for the energy $E$ in
nucleus~2 and at the channel radii.
This equation results from applying Eqs.~(58)-(60) of \textcite{Lan66};
see also Ref.~\cite{Hal90}.
If the internal matrix elements $\langle\lambda|\Delta H|\mu\rangle$ could
be evaluated, this equation could be put into standard form by diagonalization
and the $R$-matrix parameters for nucleus~2 would be determined.
In a phenomenological analysis, this avenue is unavailable.
The operator $\Delta H$ consists of Coulomb potentials and possibly
charge-symmetry violating nuclear interactions.
It is expected that the dominant contribution to
$\langle\lambda|\Delta H|\mu\rangle$ will be a constant Coulomb
energy shift along the diagonal, with variations along the diagonal
and off-diagonal elements being much smaller. I thus assume that
$\langle\lambda|\Delta H|\mu\rangle=\Delta_\lambda\delta_{\lambda\mu}$.
In this case, no diagonalization is necessary and the $R$-matrix
parameters for nucleus~2 are
\begin{subequations} \label{eq:mirror_trans}
\begin{align} \label{eq:mirror_E_shift}
\hat{E}_\lambda(2) &= \hat{E}_\lambda(1) + \Delta_\lambda \quad\quad\mbox{and} \\
\hat{\gamma}_{\lambda c}(2) &= \pm\hat{\gamma}_{\lambda c}(1) . \label{eq:gamma_1_2}
\end{align}
\end{subequations}
I will usually assume constant $\Delta_\lambda=\Delta$, but allowing the diagonal
elements to vary provides the flexibility needed to exactly match
the resonance energies in nucleus~2 to experimental values, if desired.
The sign in Eq.~(\ref{eq:gamma_1_2}) is chosen to be consistent with
Eq.~(\ref{eq:reduced_iso}), which predicts that
mirror reduced widths will at most differ by a change in sign.
Finally, the level parameters
$\{\hat{E}_\lambda(2),\hat{\gamma}_{\lambda c}(2),B_c\}$
may be transformed into resonance parameters for nucleus 2,
$\{E_i(2),\gamma_{ic}(2)\}$, using the method described in
Appendix~\ref{app:transforms}.
For $\Delta_\lambda=\Delta$, this procedure is independent of the $B_c$
values used.

These procedures produce energy shifts of the resonance levels in
mirror nuclei in addition to the $\Delta_\lambda$. The additional changes
arise from the differences in the external wave functions (i.e., coupling to
the continuum). These shifts are the multilevel generalization of the
well-known Thomas-Ehrman shift~\cite{Tho52,Ehr51}.
In addition, a particular reduced width amplitude $\gamma_{ic}(2)$
has in general a parentage in {\em all} of the $\gamma_{jc}(1)$.
This mixing leads to a breaking of the simple single-level isospin relation,
Eq.~(\ref{eq:reduced_iso}), for the resonant reduced width amplitudes.
In the multilevel case, this equation should applied instead to
the reduced width amplitudes of the states satisfying energy- and
isospin-independent boundary conditions.

\subsection{Application to \texorpdfstring{$2^+$}{2+} states in
\texorpdfstring{${}^{18}{\rm O}$}{18O} and
\texorpdfstring{${}^{18}{\rm Ne}$}{18Ne}}

The mirror nuclei ${}^{18}{\rm O}$ and ${}^{18}{\rm Ne}$ have three
$2^+$ states with significant spectroscopic strength in nucleon decay channels
that are located near the nucleon separation threshold (particularly
in the case of ${}^{18}{\rm Ne}$).
This system thus provides a good case for demonstrating the non-trivial
effects that may arise. The importance of continuum mixing in this
case has been noted and studied previously using the shell model embedded
in the continuum~\cite{Chat06,Oko12}.
The ANC of the second $2^+$ state in ${}^{18}{\rm Ne}$ plays an important
role in determining the rate of the ${}^{17}{\rm F}(p,\gamma){}^{18}{\rm Ne}$
reaction in novae~\cite{Chat06,AlA14,Kuv17}.
The notation $(1)$ and $(2)$ will often be utilized to indicate
${}^{18}{\rm O}$ and ${}^{18}{\rm Ne}$ in this subsection.

\begin{table}[tbh]
\caption{Adopted information for the first three $2^+$ states of
  ${}^{18}{\rm O}$ and ${}^{18}{\rm Ne}$.} \label{tab:A_18_exp}
\begin{tabular}{cccccc} \hline\hline \\[-2.0ex]
& \multicolumn{2}{c}{${}^{18}{\rm O}$} & \phantom{X}
  &\multicolumn{2}{c}{${}^{18}{\rm Ne}$} \\
  \cline{2-3}\cline{5-6} \rule{0pt}{1.2em}
$n\ell_j$ & $E_x$ & $C^2_c$ && $E_x$ & $C^2_c$ or $\Gamma_c$ \\
          & (keV) & (fm${}^{-1}$) && (keV) & (fm${}^{-1}$ or keV) \\
\hline \\[-2.0ex]
$2s_{1/2}$ & \multirow{2}{*}{1982.1} & $5.77\pm 0.63$ & &
            \multirow{2}{*}{1887.3} & $16.0\pm 8.0$ \\ \rule{0pt}{1.0em}
$1d_{5/2}$ & & $2.10\pm 0.23$ & & & $2.6\pm 1.2$ \\  \rule{0pt}{1.0em}
$2s_{1/2}$ & \multirow{2}{*}{3920.4} & $4.11\pm 0.62$ & &
            \multirow{2}{*}{3616.4} & $148\pm 56$ \\ \rule{0pt}{1.0em}
$1d_{5/2}$ & & $0.45\pm 0.06$ & & & $3.1\pm 1.2$ \\
$2s_{1/2}$ & \multirow{2}{*}{5254.8} & $2.18\pm 0.33$ & &
            \multirow{2}{*}{$5098\pm 8$} & $44.5\pm 1.7$ \\ \rule{0pt}{1.0em}
$1d_{5/2}$ & & $0.0080$ & & & - \\
\hline\hline
\end{tabular}
\end{table}

The available information for the excitation energies and widths
or ANCs for these states is summarized in Table~\ref{tab:A_18_exp}.
The excitation energies are very well known, with the exception
of the $2^+_3$ state of ${}^{18}{\rm Ne}$, where the value adopted is
the weighted average of $5075\pm 13$~keV~\cite{Ner81},
$5099\pm 10$~keV~\cite{Ner81}, and $5106\pm 8$~keV~\cite{Hah96},
with the error rescaled to provide a $\chi^2$ of 2.
The ${}^{17}{\rm O}+n$ ANCs for the $2^+_1$ and $2^+_2$ states of
${}^{18}{\rm O}$ are taken from \textcite{AlA14},
who performed an analysis of their
${}^{13}{\rm C}({}^{17}{\rm O},{}^{18}{\rm O}){}^{12}{\rm C}$ data and
the ${}^{17}{\rm O}(d,p){}^{18}{\rm O}$ data of \textcite{Li76}.
The $2s_{1/2}$ ANC for the $2^+_3$ state of ${}^{18}{\rm O}$ is calculated
from the spectroscopic factor and binding potential reported by
Ref.~\cite{Li76}. The value was renormalized downward by 10\%, a factor
the brought the ANCs of Ref.~\cite{Li76} into agreement with Ref.~\cite{AlA14}
for the $2^+_1$ and $2^+_2$ states.
The experimental ${}^{17}{\rm O}(d,p){}^{18}{\rm O}$ angular distribution
indicates that the $1d_{5/2}$ spectroscopic factor for the $2^+_3$ state
of ${}^{18}{\rm O}$ is very small~\cite{Li76}. This finding is supported by the
shell model calculations of \textcite{Law76}. The ANC for the state
adopted in Table~\ref{tab:A_18_exp} is based on their calculations;
setting this quantity to zero does not significantly change any of the
results reported below.
The ANCs for the $2^+_1$ and $2^+_2$ states of ${}^{18}{\rm Ne}$ are taken
from the measurements of \textcite{Kuv17}.
Note that this is a difficult radioactive ion beam
experiment with limited angular coverage.
The $2s_{1/2}$ and $1d_{5/2}$ ANCs were not independently determined;
the ratio from the mirror nucleus was assumed.
As shown in the table, the uncertainties in the ANCs are rather large and are
the result of adding the experimental statistical and systematic uncertainties
reported in Ref.~\cite{Kuv17} in quadrature.
The proton width of the $2^+_3$ state of ${}^{18}{\rm Ne}$ is determined
from the weighted average of $45\pm 5$~keV~\cite{Hah96},
$45\pm 2$~keV~\cite{Cam01}, and $42\pm 4$~keV~\cite{Jin10}.
The result of \textcite{Hah96} is a total width determination and the latter
two were extracted from fits to elastic scattering that appear to
have assumed the width is entirely due to $\ell=0$ proton emission
to the ${}^{17}{\rm F}$ ground state.
The $1d_{5/2}$ single-particle width for this state is 6.6~keV, so it
is potentially possible that this channel contributes somewhat to the
total width. However, in light of the small spectroscopic factor for this
channel in the mirror state, this is unlikely.
I also note that \textcite{Cal12}
report $\Gamma_{p'}/\Gamma_p=0.11\pm 0.04$ for a combination of the
5.10- and 5.15-MeV states of ${}^{18}{\rm Ne}$. The decay of a $2^+$ state
to a proton and the first excited state of ${}^{17}{\rm F}$ requires
$\ell=2$. Considering the additional Coulomb barrier present in this case,
this reported branch to the first excited state of
${}^{17}{\rm F}$ cannot involve the $2^+$ state of ${}^{18}{\rm Ne}$.
For these reasons, the measured proton width is assigned entirely
to the $2s_{1/2}$ channel.

\begin{table}[tbh]
\caption{The quantities $\ell$, $E_i(1)$, and  $E_i(2)$ are the nucleon orbital
angular momentum and the experimental level energies relative to the nucleon
separation thresholds in ${}^{18}{\rm O}$ and ${}^{18}{\rm Ne}$, respectively.
The final two columns provide the ANC or width predictions in
${}^{18}{\rm Ne}$, treating each pair of levels independently and using two
different methods.} \label{tab:ne18_independent}
\begin{tabular}{ccccccc}
\hline\hline \\[-2.0ex]
&& ${}^{18}{\rm O}$ && \multicolumn{3}{c}{${}^{18}{\rm Ne}$} \\
\cline{3-3} \cline{5-7} \rule{0pt}{1.2em}
&&&&& Eqs.~(\ref{eq:anc_spec}) and (\ref{eq:width_spec}) &
  Eq.~(\ref{eq:R_mirror}) \\
$\ell$ & \phantom{X} & $E_i(1)$ & \phantom{X} & $E_i(2)$ &
  $C^2_c$ or $\Gamma_c$ & $C^2_c$ or $\Gamma_c$ \\
&& (MeV) && (MeV) & (fm${}^{-1}$ or keV) & (fm${}^{-1}$ or keV) \\
\hline \\[-2.0ex]
0 && \multirow{2}{*}{$-6.062$} && \multirow{2}{*}{$-2.034$} & 14.89  & 14.87 \\
2 &&                           &&                           &  2.84  &  2.59 \\
0 && \multirow{2}{*}{$-4.124$} && \multirow{2}{*}{$-0.305$} & 117.4  & 125.2 \\
2 &&                           &&                           &  2.48  &  2.20 \\
0 && \multirow{2}{*}{$-2.790$} && \multirow{2}{*}{\phantom{$-$}1.176}
                                                            & 102    & 130 \\
2 &&                           &&                           &
  $1.9\times 10^{-2}$   & $1.7\times 10^{-2}$ \\
\hline\hline
\end{tabular}
\end{table}

\begin{figure}
\includegraphics[width=\columnwidth]{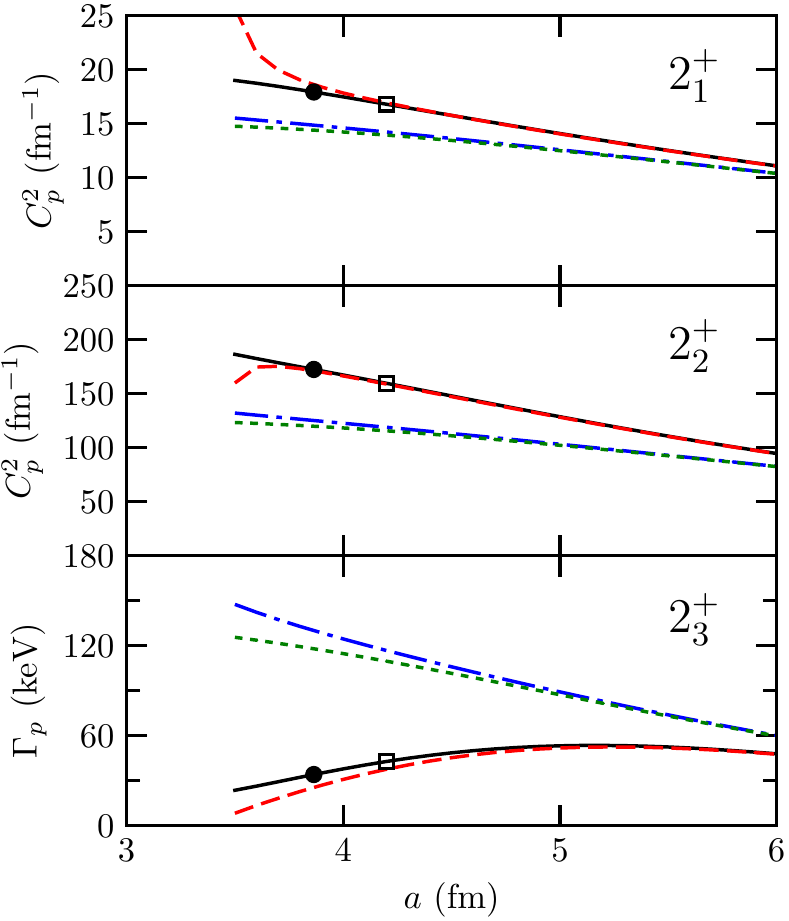}
\caption{Predicted $\ell=0$ ANCs or widths for the first three $2^+$
states in ${}^{18}{\rm Ne}$ versus channel radius for various approaches.
The results for treating the levels independently, using
Eq.~((\ref{eq:R_mirror}), are shown for the tail of the nuclear potential
included (excluded) by the blue dash-dotted (green dotted) curve.
The results for treating the levels simultaneously are shown
for the tail of the nuclear potential included (excluded)
by the solid black (red dashed) curve.
The plotted points show the results for the simultaneous treatment with
the tail of the nuclear potential included, for the preferred channel radius
of 3.86~fm used for the calculations shown in
Table~\ref{tab:multi_level_mirror} (solid circles) and for a channel radius
of 4.2~fm (open squares).}
\label{fig:o18_ne18_chan_rad}
\end{figure}

I first investigated the results of treating the levels independently, using
two different methods, as shown in Table~\ref{tab:ne18_independent}.
\textcite{AlA14} predicted ANCs in ${}^{18}{\rm Ne}$ from
the experimental values for ${}^{18}{\rm O}$, assuming the
spectroscopic factors $\mathcal{S}_c$ are the same for both members of
the mirror pair, using Eq.~(\ref{eq:anc_spec}).
I have used the same approach for the ANCs or widths [using
Eq.~(\ref{eq:width_spec})], with the results shown in the fourth column.
These findings are in good agreement with their work for the first two levels,
the only $2^+$ states analyzed in Ref.~\cite{AlA14}.
For the remainder of the calculations shown in this subsection, the
depth of the Woods-Saxon potential was fixed at 53.5~MeV, which places
the $\ell=0$ single-particle states at $-4.03$ and $-0.48$~MeV relative
to the nucleon separation thresholds in ${}^{18}{\rm O}$ and
${}^{18}{\rm Ne}$, respectively.
The fifth column shows the results of applying Eq.~(\ref{eq:R_mirror})
to determine the ANCs or widths. 
Little difference is seen, except for
the $\Gamma_p$ for the $2^+_3$ state, which is about 30\% larger in
the latter approach.
The single-particle width of this state is rather broad, about 330~keV,
which is the likely reason for some of the difference in this case.
There is also little sensitivity to the assumed nuclear potential:
Neglecting it entirely changes the results by less than 10\%, for
the preferred channel radius of 3.86~fm. This value corresponds to
$a=R_n+a_n$, as discussed in Subsec.~\ref{subsec:practical}, and
lies just outside the nuclear surface, such that additional mirror symmetry
breaking is avoided. This sensitivity to the tail
of the nuclear potential and channel radius are shown in
Fig.~\ref{fig:o18_ne18_chan_rad}. 
For the $2^+_1$ and $2^+_2$ states, the predicted ANCs in ${}^{18}{\rm Ne}$
are in good agreement with the experimental values of \textcite{Kuv17}
shown in Table~\ref{tab:A_18_exp}, although the large experimental errors
preclude any accurate statement.
However, for the $2^+_3$ state, the predicted proton widths are more than
a factor of two larger than the accurately-known experimental value.
The calculations using Eq.~(\ref{eq:R_mirror}) included the $\ell=2^*$
channel introduced below, using the $\gamma_{ic}(1)$ from
Table~\ref{tab:multi_level_mirror}. This consideration had very
little effect.

\begin{table*}[tbh]
\caption{Transformation of the ${}^{18}{\rm O}$ (1) resonance parameters
to ${}^{18}{\rm Ne}$ (2) resonance parameters. The meaning of the various
quantities is described in the text. The final column gives the resulting
ANCs or widths in ${}^{18}{\rm Ne}$.}
\label{tab:multi_level_mirror}
\begin{tabular}{@{\extracolsep{0.2in}}ccccccccc}
$\ell$ & $E_i(1)$ & $\gamma_{ic}(1)$ & $\hat{E}_\lambda(1)$ &
  $\hat{\gamma}_{\lambda c}$ & $\hat{E}_{\lambda}(2)$ &
  $E_i(2)$ &  $\gamma_{ic}(2)$ & $C^2_c$ or $\Gamma_c$ (2) \\
\hline\hline \\[-2.0ex]
& (MeV) & (MeV${}^{1/2}$) & (MeV) & (MeV${}^{1/2}$) & (MeV) & (MeV) &
  (MeV${}^{1/2}$) & (fm${}^{-1}$ or keV)  \\
\hline \\[-2.0ex]
0\phantom{${}^*$} & $-6.062$ & $-0.747$ & $-8.099$ & $-0.999$ & $-3.322$ &
  $-2.023$ & $-0.835$ & 17.94 \\
2\phantom{${}^*$} & & $-1.483$ & & $-1.420$ & & & $-1.494$ & 2.53 \\
$2^*$ & & $-0.609$ & & $-0.814$ & & & $-0.681$ & 0.68 \\ \hline \\[-2.0ex]
0\phantom{${}^*$} & $-4.124$ & \phantom{$-$}0.896 & $-5.083$ &
  \phantom{$-$}1.158 & $-0.305$ & $-0.305$ &
  \phantom{$-$}1.158 & 172.5 \\
2\phantom{${}^*$} & & $-1.185$ & & $-1.343$ & & & $-1.343$ & 2.34 \\
$2^*$ & & \phantom{$-$}0.730 & & \phantom{$-$}0.944 & & & \phantom{$-$}0.944 &
  0.48 \\ \hline \\[-2.0ex]
0\phantom{${}^*$} & $-2.790$ & $-0.789$ & $-2.798$ &
  $-0.782$ & \phantom{$-$}1.979 & \phantom{$-$}1.594 &
  $-0.343$ & 34.2 \\
2\phantom{${}^*$} & & \phantom{$-$}0.075 & & \phantom{$-$}0.075 & & &
  \phantom{$-$}0.113 & $5.2\times10^{-2}$ \\
$2^*$ & & $-0.643$ & & $-0.637$ & & & $-0.280$ & $1.6\times10^{-2}$ \\
\hline\hline
\end{tabular}
\end{table*}

In this case, there are three important channels with thresholds located in
the neighborhood of the first three $2^+$ levels.
The first two are the $\ell=0$ and $\ell=2$ $n+{}^{17}{\rm O}$ or
$p+{}^{17}{\rm F}$ channels already discussed.
The nuclei ${}^{17}{\rm O}$ and ${}^{17}{\rm F}$ have low-lying
$1/2^+$ states, thus leading to an additional important channel and
couples $\ell=2$ nucleons to these excited states.
These channels will be indicated as $\ell=2^*$.
No experimental information for these channels is available.
In addition, the relative signs of the reduced width amplitudes within
a particular channel have a significant impact in the
transformation process described in Appendix~\ref{app:transforms}.
These unknown parameters can be fixed using the shell model.
In a simple shell model picture for the $2^+$ states, with two $T=1$ nucleons
outside an ${}^{16}{\rm O}$ core, both the $\ell=0$ and $\ell=2^*$
channels arise from the $(d_{5/2},s_{1/2})$ component of the wave function.
This consideration leads to the spectroscopic amplitudes being equal,
up to an overall sign that is irrelevant in the present case.
More detailed calculations were performed using the code
{\sc nushellx}~\cite{Bro14}, with the Zuker-Buck-McGrory (ZBM) model space
and interactions for nucleons outside a ${}^{12}{\rm C}$ core~\cite{Zuk68}.
Both interactions given by ZBM were utilized.
For the first three $2^+$ states, the ratio of the $\ell=2^*$ to
$\ell=0$ spectroscopic amplitude was always found to lie between
0.96 and 0.88, depending somewhat upon the particular state and interaction.
The simple picture is thus confirmed within a good degree of accuracy
and will be used below to estimate the parameters for the $\ell=2^*$ channels.
This calculation also predicts the signs for all of the channels,
where good agreement is seen for both interactions and also the three
calculations of \textcite[Table~IV]{Law76}.
The lone exception is for the $\ell=2$ channel for the $2^+_3$ state,
where the spectroscopic amplitude is very small.
In this case, the sign from Ref.~\cite{Law76} is utilized, although
it has no impact on the results reported below.

The three states are then treated simultaneously as described in
Subsec.~\ref{subsec:gen_phen_mirror}. The reduced width amplitudes
$\gamma_{ic}(1)$ in ${}^{18}{\rm O}$ were determined from the experimental
${}^{18}{\rm O}$ ANCs in Table~\ref{tab:A_18_exp} using
Eq.~(\ref{eq:gamma_c_x_c}).
The relative signs of the reduced width amplitudes within a particular
channel have a significant impact on the
transformation process described in Appendix~\ref{app:transforms}.
These signs are taken from the shell model calculations described above.
Further following the shell model, the $\ell=2^*$ reduced-width amplitudes
were adjusted such that
$\hat{\gamma}_{\lambda,\ell=2^*}=0.815\hat{\gamma}_{\lambda, \ell=0}$,
where $0.815$ is the ratio of single-particle reduced-width amplitudes.
The boundary-condition constants $B_c$ were chosen to equal to the
shift function for the $2^+_2$ state in ${}^{18}{\rm Ne}$.
The parameters $\{E_i(1),\gamma_{ic}(1)\}$ are then transformed
to $\{\hat{E}_{\lambda c}(1),\hat{\gamma}_{\lambda c}\,B_c\}$.
A constant shift $\Delta=4.777$~MeV was used in Eq.~(\ref{eq:mirror_E_shift}),
to match exactly $E_{2}(2)$ to the experimental energy of the $2^+_2$ state.
The $\hat{\gamma}_{\lambda c}$ do not change sign for this mirror transformation.
Then the parameters  $\{\hat{E}_{\lambda c}(2),\hat{\gamma}_{\lambda c}\,B_c\}$
are transformed to $\{E_i(2),\gamma_{ic}(2)\}$,
the resonance parameters in ${}^{18}{\rm Ne}$.
Finally, the ANCs or widths are calculated from the resonance parameters
using Eq.~(\ref{eq:x_c_gamma_c}). Since the calculated $E_i(2)$ do not
necessarily exactly match the experimental values, the experimental
energy values are used in this last step.
The resulting ANCs or widths and the parameter values at the steps of this
process are shown in Table~\ref{tab:multi_level_mirror}.

Significant differences are seen compared to the results considering
each level independently. The squares of the ANCs of the first two
$2^+$ states in
${}^{18}{\rm Ne}$ are predicted to be significantly larger by the
multi-level calculation. This result is still in agreement with the
experimental result, due to the large experimental error.
In addition, the width of the $2^+_3$ state
is found to be about a factor of three smaller, such that the prediction
is now below the experimental value.
These finding were found to be sensitive to several ingredients
in the calculation.
The dominant sensitivity is to the the $\ell=0$ reduced-width amplitudes
and their signs, but the $\ell=2$ and $\ell=2^*$ channels also
contribute non-trivially.
Changing the sign of the input $\ell=0$ $2^+_2$ reduced-width amplitude
causes the predicted
ANC of the $2^+_2$ state in ${}^{18}{\rm Ne}$ to be {\em smaller}
than that found when the levels are considered independently.
It also is found that all of the first three $2^+$ states play an important
role in this mixing.

Several other factors were investigated that had little influence
on these results. A background level with physically reasonable
reduced-width parameters placed at $E_4=10$~MeV in
${}^{18}{\rm O}$ was found to have little effect. Neglecting the
tail of the nuclear potential was likewise found to have little effect,
as shown in Fig.~\ref{fig:o18_ne18_chan_rad}.

The energy shift utilized leads to $E_1(2)$ being overpredicted by 12~keV
and $E_3(2)$ being overpredicted by 418~keV, compared to the experimental
values. Differences of up to a few hundred keV are expected, because the
actual Coulomb energy shift includes contributions that depend upon the
specific internal structure of the state~\cite{Ner81,She98}.
The constant energy shift was varied to match
the energies of the $2^+_1$ and $2^+_3$ states, and the predicted
ANCs and widths were found to not change significantly. Calculations were also
performed using level-dependent shifts in Eq.~(\ref{eq:mirror_E_shift})
that allowed all of the $E_i(2)$ to match experiment values.
Again, no significant changes in the predicted ANCs or widths were found.
In this case, the results become slightly $B_c$ dependent, as discussed
in Appendix~\ref{app:transforms}. These calculations were done using various
$B_c$ values, including values matching the shift function for other
levels and $B_c=0$. None made a significant difference in the predictions.

The dependence of the calculation on the channel radius and the tail of
the nuclear potential is shown in Fig.~\ref{fig:o18_ne18_chan_rad}, along
with calculations treating the levels separately.
All calculations converge to the same result for large channel radii, as
expected since more of the wave functions are inside the channel radii.
Mathematically, the volume renormalization factors approach unity and
the shift factors approach zero in this limit.
However, the most physically correct channel radius is one just outside
the nuclear surface, as discussed in Subsec.~\ref{subsec:overlap_phenom}.

There are some indications that this calculation overpredicts the
mixing effects. The width of the $2^+_3$ state is overcorrected, with
the predicted value of 34.2~keV being about 30\%
{\em below} the experimental value.
Also, the energy of this state is overpredicted by 421~keV, which
is more than expected from differences in the internal Coulomb energy.
These issues could be due to channel nonorthogonality, as mentioned in
Subsec.~\ref{subsec:overlap_phenom}, since this case has three
channels with significant spectroscopic strength.
If the channel radius is modestly increased to 4.2~fm, these discrepancies
with experiment are much reduced, with the overpredictions
of $E_1(2)$ and $E_3(2)$ becoming 28 and 169~keV, respectively, and
the prediction for the width of the $2^+_3$ state becoming 42.9~keV.
The predictions for the $\ell=0$ $C^2_p$ for the $2^+_1$ and
$2^+_2$ states are then 16.81 and 159.7~fm${}^{-1}$, respectively.
The predictions for channel radii of 3.86 and 4.2~fm are indicated
in Fig.~\ref{fig:o18_ne18_chan_rad}.

This model of external mixing correctly predicts the striking reduction
by a factor of 2-3 in the predicted width of the 
$2^+_3$ state in ${}^{18}{\rm Ne}$ compared to using naive mirror symmetry.
Take the average of the $a=3.86$ and 4.2~fm results, I recommend
\begin{equation}
\begin{split}
& C^2_p(2^+_1)=17.4\pm 2.6~\mathrm{fm}^{-1} \quad\mbox{and} \\
& C^2_p(2^+_2)=166\pm 25~\mathrm{fm}^{-1} ,
\end{split}
\end{equation}
for the $\ell=0$ ANCs of the first two $2^+$ states of ${}^{18}{\rm Ne}$,
using mirror symmetry.
The 15\% uncertainty is estimated from
the various model uncertainties discussed above; the experimental errors
on the input mirror ANCs given in Table~\ref{tab:A_18_exp}
contribute an additional 15\% uncertainty.
The value for the $2^+_2$ state is 42\% higher than the result of
\textcite{AlA14}, that was extracted using naive mirror symmetry.
This result is 12\% higher than the determination of
\textcite{Kuv17} that does not rely upon mirror symmetry, but this
difference is well within their 35\% uncertainty.
The present result would lead to a somewhat higher reaction rate for
${}^{17}{\rm F}(p,\gamma){}^{18}{\rm Ne}$ in novae.
A re-evaluation of this rate will not be attempted here.
At this time, one is placed in the difficult position of choosing between
using the more accurate information available from the mirror nucleus
or the less accurate measurements in ${}^{18}{\rm Ne}$.
An improved experimental determination
of the $2^+$ ANCs in ${}^{18}{\rm Ne}$ would be most helpful here.

The importance of external mixing has been noted previously in
this case~\cite{Chat06,Tim08,Oko12}. \textcite{Tim08} performed three-body
calculations considering either two neutrons or two protons outside an
inert ${}^{16}{\rm O}$ core. They report smaller mirror symmetry breaking
effects than reported here. However, it is known that four-particle--two-hole
excitations (i.e., excitations of the ${}^{16}{\rm O}$ core) must be
taken into account in order to describe the first three
$2^+$ states~\cite{Law76}, making this difference unsurprising.
Calculations using the shell model embedded in the continuum have been
reported by \textcite{Oko12}. For the $2^+_1$ state they find an
increase in the ratio of the ${}^{18}{\rm Ne}$ to ${}^{18}{\rm O}$ ANCs that is
similar to this work.
For the $2^+_2$ state, they report a decrease in this ratio, in the
{\em opposite} direction of the significant increase found here.
They did not report results for the $2^+_3$ state.
It appears that much of this difference can be attributed to the
present calculation being tuned to experimental ANCs in ${}^{18}{\rm O}$.
For example, they report~\cite[Table~VII]{Oko12} squared $\ell=0$ ANCs for the
$2^+_2$ state of ${}^{18}{\rm O}$ that are 50-85\% larger than
the experimental value.
If their ${}^{18}{\rm O}$ ANCs for the first two states are used in
the present calculations, the discrepancy largely goes away.
However, their results for the $2^+_3$ state would need be included in order
to make a definitive comparison of the two approaches.

\subsection{Discussion}

If the off-diagonal components of the matrices in
Eqs.~(\ref{eq:mmatrix}), (\ref{eq:nmatrix}), and~(\ref{eq:non_linear_eigen})
are zero, the transformation process becomes identical to treating the
levels independently.
This situation would occur if the shift factors were independent of
energy and the boundary condition constants were taken equal to
these shift factors.
It can thus be said that external mixing is driven by the energy
dependence of the outgoing-wave boundary condition.
However, from this discussion in Sec.~\ref{sec:single_particle}, this
energy dependence is intimately related to the extension of wave functions
beyond the channel radius.
This quantity is largest near separation thresholds and for
low orbital angular momentum.
Note also that the energy dependence of the shift factor also gives rise
to the volume renormalization factor.
The magnitude of the off-diagonal elements is also proportional
to the reduced-width amplitudes.
The symmetry breaking for the resonant states then results because the
ingredients listed above are modified for mirror states by
the different charges and separation energies.
The $2^+_2$ and $2^+_3$ states of ${}^{18}{\rm O}$ and ${}^{18}{\rm Ne}$
are thus ideal for exposing this phenomenon since they couple with
significant spectroscopic strength to nucleons with $\ell=0$ and
are located near the nucleon separation threshold.

\section{Conclusions}

This work reviewed the relationship between spectroscopic factors
and single-particle wave functions and their physical counterparts,
ANCs and widths. $R$-matrix theory was used extensively to
describe these relationships. Also, particular attention was paid to
effects arising from beyond the channel radii, which may be termed
coupling to the continuum. These effects may be large for levels
near a channel threshold, if the level couples significantly to that channel.

A natural application of these concepts is isospin and mirror symmetry.
$R$-matrix theory is an efficient tool to study the symmetry breaking
in analog or mirror states arising from differences in the wave functions
beyond the channel radii.
The examples of single levels in nucleon~+~${}^{12}{\rm C}$,
nucleon~+~${}^{16}{\rm O}$, and nucleon~+~${}^{26}{\rm Al}$ were studied.
It is straightforward to extend this analysis to a group of levels, in
which case the continuum coupling may cause a mixing of the levels.
The first three $2^+$ states of ${}^{18}{\rm O}$ and ${}^{18}{\rm Ne}$
were studied in this manner. It was found that the ANC of the second
$2^+$ state in ${}^{18}{\rm Ne}$ deduced from the mirror state in
${}^{18}{\rm O}$ is significantly larger than found in previous work.
This finding has the effect of increasing the
${}^{17}{\rm F}(p,\gamma){}^{18}{\rm Ne}$ reaction rate in novae.

The concepts described in this paper arise frequently in the analysis of
transfer reaction experiments, the use of theoretical spectroscopic factors
to determine ANCs or widths, and the prediction of ANCs or widths using
mirror symmetry. It is hoped that this paper will allow future analyses
of this type to be carried out with greater confidence and clarity.

\begin{acknowledgments}
The impetus for this work was provided in part by the author's participation
in {\em TALENT Course 6: Theory for Exploring Nuclear Reaction Experiments},
organized by the FRIB Theory Alliance in June 2019.
I thank Gerry Hale for contributing to the formulation
of the algorithm presented in Appendix~\ref{app:dlde}.
I thank Bing Guo, Gavin Lotay, and Steve Pain for helpful
discussions regarding some of the experimental results analyzed in this paper.
I also thank James deBoer for useful comments on the manuscript.
This work was supported in part by the U.S. Department of Energy,
under Grants No.~DE-FG02-88ER40387 and No.~DE-NA0003883.
\end{acknowledgments}

\appendix

\section{Algorithm for computing
  \texorpdfstring{$\protect\bm{\partial L/\partial E}$}{dL/dE} }
\label{app:dlde}

\begin{table*}[tbh]
\caption{Starting values ($n=1$) and recurrence formulas ($n>1$)
for the sequences to evaluate $L$ and $\dot{L}$.}
\label{tab:steed}
\begin{tabular}{rcl@{\hspace*{0.5in}}rcl} \hline\hline \\[-2.0ex]
\multicolumn{3}{c}{$L$} & \multicolumn{3}{c}{$\dot{L}$} \\ \hline \\[-2.0ex]
\multicolumn{6}{c}{starting values\hspace*{1.0in}} \\[1ex]
$a_1$ &$=$& $-\eta^2-\ell(\ell+1)+i\eta$ &
     $\dot{a}_1$ &$=$& $-2\eta+i$ \\[1ex]
$b_1$ &$=$& $2(\rho-\eta+i)$ &
     $\dot{b}_1$ &$=$& $-2$ \\[1ex]
$D_1$ &$=$& $1/b_1$ &
     $\dot{D}_1$ &$=$& $-\dot{b}_1/b_1^2=2/b_1^2$ \\[1ex]
$\Delta h_1$ &$=$& $i a_1 D_1$ &
     $\Delta\dot{h}_1$ &$=$& $i(\dot{a}_1D_1+a_1\dot{D}_1)$ \\[1ex]
$h_1$ &$=$& $i(\rho-\eta)+\Delta h_1$ &
     $\dot{h}_1$ &$=$& $-i+\Delta\dot{h}_1$ \\ \hline \\[-2.0ex]
\multicolumn{6}{c}{recurrence formulas\hspace*{1.0in}} \\[1ex]
$a_n$ &$=$& $a_{n-1} +2(n-1) + 2i\eta$ &
     $\dot{a}_n$ &$=$& $\dot{a}_{n-1} + 2i$ \\[1ex]
$b_n$ &$=$& $b_{n-1} + 2i$ &
     $\dot{b}_n$ &$=$& $\dot{b}_{n-1}=\dot{b}_1=-2$ \\[1ex]
$D_n$ &$=$& $(D_{n-1} a_n + b_n)^{-1}$ &
     $\dot{D}_n$ &$=$& $-\dfrac{\dot{D}_{n-1}a_n+D_{n-1}\dot{a}_n+\dot{b}_n}%
     {(D_{n-1}a_n+b_n)^2}$ \\[2ex]
$\Delta h_n$ &$=$& $(b_n D_n -1)\Delta h_{n-1}$ &
     $\Delta\dot{h}_n$ &$=$&
     $(\dot{b}_nD_n+b_n\dot{D}_n)\Delta h_{n-1}+(b_nD_n-1)\Delta\dot{h}_{n-1}$
     \\[1ex]
$h_n$ &$=$& $h_{n-1} + \Delta h_n$ &
     $\dot{h}_n$ &$=$& $\dot{h}_{n-1}+\Delta\dot{h}_n$ \\
\hline\hline
\end{tabular}
\end{table*}

The quantity $\partial L/\partial E$, where the derivative is taken with
fixed radius and $L$ is the logarithmic radial derivative
of the outgoing Coulomb wave defined by Eq.~(\ref{eq:L_outgoing}),
is important for normalizing bound and/or Gamow states, as well
as relating observed and reduced widths in $R$-matrix calculations.
Existing methods for calculating this quantity include the numerical
differentiation of $L$ values calculated using standard Coulomb function
routines and numerical quadrature~\cite{Ver82}.
For the uncharged case, an analytic result is available; see
\textcite[Eq.~(6)]{Gya71} and Eqs.~(\ref{eq:l_prime})
and~(\ref{eq:dlde_via_eta_rho}) below.
Here, I present a more efficient and more accurate approach to
computing $\partial L/\partial E$ for the general Coulomb case that can be
performed in parallel with the calculation of $L$ itself.

Modern numerical routines for the computation of Coulomb
wave functions~\cite{Tho85,Tho86,Mic07} use a continued fraction
technique to calculate $L$ as an intermediate step for much
of the $\ell$-$\rho$-$\eta$ parameter space.
The continued-fraction algorithm is described in detail
by \textcite{Bar74}, and I will utilize their notation and work in
terms of the dimensionless variables $\rho$ and $\eta$.
The approach is to apply the energy derivative to the continued fraction
analytically.
The derivative $\partial E$ is to be evaluated at fixed radius,
implying $\rho\eta$ is constant.
Using $\partial\rho/\partial E=\rho/2E$, one finds
\begin{equation}
\frac{\partial}{\partial E}=\frac{\rho}{2E}\left(
\frac{\partial}{\partial \rho}-\frac{\eta}{\rho}\frac{\partial}{\eta}\right) ,
\end{equation}
when $\rho$ and $\eta$ are considered independent variables.
The outgoing Coulomb wave $O$ satisfies
\begin{equation}
O^{\prime\prime} +\left[1-\frac{2\eta}{\rho}-\frac{\ell(\ell+1)}{\rho^2}\right]O=0,
\end{equation}
where ${}^\prime\equiv d/d\rho$. Since $L=\rho O^\prime/O$, one has
\begin{equation}
L^\prime = \frac{O^\prime}{O}+\rho\left[\frac{O^{\prime\prime}}{O}-
\left(\frac{O^\prime}{O}\right)^2\right]
\end{equation}
and hence
\begin{equation} \label{eq:l_prime}
L^\prime=\frac{1}{\rho}\left[L(1-L)+\ell(\ell+1)\right]+2\eta-\rho.
\end{equation}
Defining $\dot{}\equiv \partial/\partial\eta$, one then has
\begin{equation} \label{eq:dlde_via_eta_rho}
\frac{\partial L}{\partial E} = \frac{\rho}{2E}\left( L^\prime -
\frac{\eta}{\rho}\dot{L}\right).
\end{equation}
Note that when $\eta=0$ this equation provides an analytic result that
can be expressed in terms of spherical Hankel functions.
If $\eta=0$ and $\ell$ is an integer, then $L$ and $L^\prime$ are rational
functions of $\rho$ and the infinite sequence for $L$ given below terminates.

Steed's algorithm~\cite[Eq.~(32)]{Bar74} provides a sequence of
$\Delta h_n$ and $h_n$ values with
\begin{equation} \label{eq:hn}
h_n = \left\{\begin{array}{l@{\hspace*{0.2in}}l}
  i(\rho-\eta) & n=0 \\
  h_0 + \sum_{k=1}^n \Delta h_k & n>0
\end{array} \right.
\end{equation}
such that $\displaystyle{\lim_{n\to\infty}} h_n=L$.
The starting values and recurrence formulas for the $\Delta h_n$
sequence are given in the first column of Table~\ref{tab:steed}.
Differentiating Eq.~(\ref{eq:hn}) with respect to $\eta$ yields
\begin{equation} \label{eq:hn_dot}
\dot{h}_n = \left\{\begin{array}{l@{\hspace*{0.2in}}l}
  -i & n=0 \\
  \dot{h}_0 + \sum_{k=1}^n \Delta\dot{h}_k & n>0
\end{array} \right. .
\end{equation}
Assuming that the sum can be differentiated term by term in the limit that
$n\to\infty$, one then has
\begin{equation} \label{eq:L_limit}
\dot{L} = \lim_{n\to\infty} \dot{h}_n .
\end{equation}
The starting values and recurrence formulas for the $\Delta\dot{h}_n$
sequence are straightforward to calculate by differentiation and
are given in the second column of Table~\ref{tab:steed}.
Note that $L$ and $\dot{L}$ are calculated
in parallel, as the $\dot{L}$ sequence depends upon the $L$ sequence.
With $L$ and $\dot{L}$ in hand, $\partial L/\partial E$ may be calculated
using Eqs.~(\ref{eq:l_prime}) and~(\ref{eq:dlde_via_eta_rho}).

\begin{table*}
\caption{The number of iterations required to reach a specified convergence
(see text) are given by $N(L)$ and $N(\dot{L})$ for $L$ and $\dot{L}$,
respectively, for some applications and corresponding values of
$\ell$, $\rho$, and $\eta$.}
\label{tab:convergence}
\begin{tabular}{@{\extracolsep{0.2in}}lccrrl} \hline\hline \\[-2.0ex]
$\ell$ & $\rho$ & \multicolumn{1}{c}{$\eta$} &
  $N(L)$ & $N(\dot{L})$ & \multicolumn{1}{c}{application} \\ \hline \\[-2.0ex]
0 & $0.480-i0.011$ & $1.404+i0.031$ & 147~ & 157~ &
  $420.5-i18.75$~keV $p+{}^{12}{\rm C}$, $r=3.51$~fm \\
0 & $2.735-i0.061$ & $1.404+i0.031$ & 35~ & 37~ &
  $420.5-i18.75$~keV $p+{}^{12}{\rm C}$, $r=20$~fm \\
0 & $i0.267$ & $-i4.261$ & 136~ & 150~ &
  $-105.2$~keV $p+{}^{17}{\rm F}$, $r=3.86$~fm \\
0 & $i1.384$ & $-i4.261$ & 39~ & 44~ &
  $-105.2$~keV $p+{}^{17}{\rm F}$, $r=20$~fm \\
0 & 0.334 & 5.662 & 238~ & 265~ &
  126.8~keV $p+{}^{26}{\rm Al}$, $r=4.35$~fm \\
0 & 1.534 & 5.662 & 81~ & 90~ &
  126.8~keV $p+{}^{26}{\rm Al}$, $r=20$~fm \\ \hline\hline
\end{tabular}
\end{table*}

A rigorous proof of Eq.~(\ref{eq:L_limit}) requires showing that the
limit of the right-hand side of the equation converges uniformly
in $\eta$ to its limit, which I have not attempted.
In practice, the sequence converges in a manner very similar
to the $h_n$ sequence.
Table~\ref{tab:convergence} shows the convergence properties for
some of the cases encountered in this work.
The quantities $N(L)$ and $N(\dot{L})$ are the $n$ values required to achieve
$|\Delta h_n/h_n|<10^{-13}$ and
$|\Delta \dot{h}_n/\dot{h}_n|<10^{-13}$, respectively.
The $\dot{h}_n$ sequence is seen to converge with just a modest number
of additional iterations compared to the $h_n$ sequence in every case.

\section{{\em R}-matrix parameter transformations}
\label{app:transforms}

Methods for transforming between $R$-matrix eigenfunctions satisfying
resonance boundary conditions for all energy levels and a basis
satisfying energy-independent boundary conditions have been given
by \textcite{Bru02}. The $N$ eigenfunctions satisfying resonance boundary
conditions correspond to level energy and reduced width parameters
$E_i$ and $\gamma_{ic}$, where $i$ is the level
index and $c$ is the channel index.
The parameters corresponding to energy-independent boundary conditions,
the assumption of traditional $R$-matrix theory~\cite{Wig47,Lan58}, are
indicated by $\hat{E}_\lambda$ and $\hat{\gamma}_{\lambda c}$.
In addition, the boundary condition parameters $B_c$ are assumed to be
real and independent of energy and isospin.
Note that the present notation differs from that of Ref.~\cite{Bru02}.

I first consider the transformation
$\{E_i,\gamma_{ic}\}\rightarrow\{\hat{E}_\lambda,\hat{\gamma}_{\lambda c},B_c\}$.
The matrices ${\bm M}$ and ${\bm N}$ are defined with elements given by
\begin{equation}
M_{ij}=\left\{
\begin{array}{ll} 1 & i=j \\ -\sum_c\gamma_{ic}\gamma_{jc}
\frac{\hat{\mathcal{S}}_{ic}-\hat{\mathcal{S}}_{jc}}{E_i-E_j} & i\neq j
\end{array} \right.
\label{eq:mmatrix}
\end{equation}
and
\begin{equation}
N_{ij} = \left\{
\begin{array}{ll} E_i+\sum_c \gamma_{ic}^2(\hat{\mathcal{S}}_{ic}-B_c) & i=j \\
\sum_c \gamma_{ic}\gamma_{jc}
\left(\frac{E_i\hat{\mathcal{S}}_{jc}-E_j\hat{\mathcal{S}}_{ic}}{E_i-E_j}
-B_c \right) & i\neq j \end{array} \right. ,
\label{eq:nmatrix}
\end{equation}
where the notation $\hat{\mathcal{S}}_{ic}$ indicates the shift function
evaluated at $E_i$.
Next one solves the real symmetric generalized linear eigenvalue equation
\begin{equation}
(\bm{N}-\hat{E}_\lambda \bm{M})\bm{b}_\lambda =0.
\end{equation}
As discussed in Ref.~\cite{Bru02}, it is expected that $\bm{M}$ is
positive definite for physically-reasonable parameters and the eigenvalue
problem can be solved to yield $N$ real eigenvalues and eigenvectors.
The eigenvectors $\bm{b}_\lambda$ may be arranged into a square matrix
$\bm{b}$ and are normalized such that $\bm{b}^T\bm{M}\bm{b}=\bm{1}$,
where $\bm{1}$ is the unit matrix.
The matrix $\bm{N}$ is also diagonalized by $\bm{b}$, with
$\bm{b}^T\bm{N}\bm{b}=\bm{e}$, where
$e_{\lambda\mu}=\hat{E}_\lambda\delta_{\lambda\mu}$.
The reduced widths $\gamma_{ic}$ and $\hat{\gamma}_{\lambda c}$
may be arranged into column matrices
$\bm{\gamma}_c$ and $\hat{\bm{\gamma}}_c$ that allow the
transformed reduced widths to be
written as $\hat{\bm{\gamma}}_c=\bm{b}^T\bm{\gamma}_c$.
This completes the transformation to the
$\{\hat{E}_\lambda,\hat{\gamma}_{\lambda c},B_c\}$ basis.

The transformation in the other direction,
$\{\hat{E}_\lambda,\hat{\gamma}_{\lambda c},B_c\}\rightarrow\{E_i,\gamma_{ic}\}$,
is accomplished by solving the real symmetric non-linear eigenvalue equation
\begin{equation} \label{eq:non_linear_eigen}
\left\{ {\bm e}-E_i{\bm 1}-\sum_c
  \hat{\bm \gamma}_c[\hat{\mathcal{S}}_c(E_i)-B_c]
  \hat{\bm \gamma}_c^T \right\} \, \bm{a}_i = 0
\end{equation}
for eigenvalues $E_i$ and eigenvectors $\bm{a}_i$.
As discussed in Ref.~\cite{Bru02}, this equation has $N$ real eigenvalues
if $\partial\hat{\mathcal{S}}_c/\partial E\ge 0$.
This condition is always met when the potential outside the channel radius
consists of the repulsive Coulomb and angular momentum barriers~\cite{Bru18}.
The tail of the attractive nuclear potential included in the calculations
presented here could spoil this situation, but in this work the
derivative has been found to be positive, for the potential strengths and
energy ranges considered.
This question would need to be revisited for $\ell=0$ neutron channels
with positive energy, where there is no Coulomb or angular momentum barrier
and any attractive potential will likely create a negative energy derivative.
The solution of non-linear eigenvalue problems has been reviewed
by \textcite{Vos14}. If the energy derivative of the shift function
is positive, the eigenvalue problem is characterized as overdamped,
which provides several nice mathematical properties~\cite{Rog64,Vos14},
including the existence of $N$ real eigenvalues noted above.
For this work, I have solved the eigenvalue equation using the
{\em safeguarded iteration} algorithm~\cite{Vos14}.
The eigenvectors are normalized such that $\bm{a}_i^T\bm{a}_i=1$ and
$\gamma_{ic}=\bm{a}_i^T\hat{\bm{\gamma}}_c$, completing the transformation.

The mirror transformation 
$\{E_i(1),\gamma_{ic}(1)\}\rightarrow\{E_i(2),\gamma_{ic}(2)\}$
is implemented as follows,
where $(1)$ and $(2)$ indicate the initial and final nuclei.
First, the resonance parameters $\{E_i(1),\gamma_{ic}(1)\}$ are transformed
to $\{\hat{E}_\lambda(1),\hat{\gamma}_{\lambda c}(1),B_c\}$.
In this basis, the boundary conditions satisfied by the eigenfunctions
are independent of isospin.
The transformation is then applied using Eq.~(\ref{eq:mirror_trans}).
Finally, one transforms
$\{\hat{E}_\lambda(2),\hat{\gamma}_{\lambda c}(2),B_c\}\rightarrow
\{E_i(2),\gamma_{ic}(2)\}$.

When not considering mirror symmetry, $\{E_i,\gamma_{ic}\}$ and physical
observables are independent of the $B_c$, even when the
number of levels in finite~\cite{Mor72,Bar72,Bru02}.
The question of $B_c$ invariance under the mirror transformation
is investigated as follows.
Equation~(\ref{eq:non_linear_eigen}) becomes
\begin{equation} \label{eq:non_linear_eigen2}
\begin{split}
& \biggl\{  {\bm e}(1)+\bm{\Delta}-E_i(2){\bm 1} \\
  & -\sum_c
  \hat{\bm \gamma}_c(1)[\hat{\mathcal{S}}_c(E_i(2))-B_c]
  \hat{\bm \gamma}_c^T(1) \biggr\} \, \bm{a}_i = 0 ,
\end{split}
\end{equation}
where the components of $\bm{\Delta}$ are given by
$\Delta_\lambda\delta_{\lambda\mu}$.
Note also that the charge used to evaluate the shift function must also
change when $1\rightarrow 2$.
The transformation to different boundary conditions,
$\{\hat{E}_\lambda (1),\hat{\gamma}_{\lambda c}(1),B_c\}\rightarrow
\{\hat{E}_\lambda '(1),\hat{\gamma}_{\lambda c}'(1),B_c'\}$,
is given by~\cite{Bar72}
\begin{equation}
\bm{e}'(1) = \bm{K}\bm{C}\bm{K}^T \quad\mbox{and}\quad
\hat{\bm{\gamma}}_c'(1) = \bm{K} \hat{\bm{\gamma}}_c(1) ,
\end{equation}
where the real orthogonal matrix $\bm{K}$ diagonalizes
\begin{equation}
\bm{C} = \bm{e}(1)-\sum_c \hat{\bm{\gamma}}_c(1)(B_c'-B_c)
  \hat{\bm{\gamma}}_c^T(1) .
\end{equation}
With $\bm{a}_i'=\bm{K}\bm{a}_i$, Eq.~(\ref{eq:non_linear_eigen2}) becomes
\begin{equation} \label{eq:non_linear_eigen3}
\begin{split}
  & \biggl\{  {\bm e}'(1) +\bm{K}\bm{\Delta}\bm{K}^T-E_i(2){\bm 1} \\
  & -\sum_c
  \hat{\bm \gamma}_c '(1)[\hat{\mathcal{S}}_c(E_i(2))-B_c']
  \hat{\bm \gamma}_c'^T (1) \biggr\} \, \bm{a}_i' = 0 .
\end{split}
\end{equation}
If $[\bm{K},\bm{\Delta}]=0$, then this equation becomes
\begin{equation} \label{eq:non_linear_eigen4}
\begin{split}
  & \biggl\{  {\bm e}'(1) +\bm{\Delta}-E_i(2){\bm 1} \\
  & -\sum_c \hat{\bm \gamma}_c '(1)[\hat{{S}}_c(E_i(2))-B_c']
  \hat{\bm \gamma}_c'^T (1) \biggr\} \, \bm{a}_i' = 0 ,
\end{split}
\end{equation}
which is of the same form as Eq.~(\ref{eq:non_linear_eigen2}) and
has the same energy shifts.
The two equations are related by the similarity transformation $\bm{K}$.
In this case, the eigenvalues
$E_i(2)$ and reduced widths $\bm{\gamma}_{ic}(2)=
\bm{a}_i^T\hat{\bm{\gamma}}_c(2)=\bm{a}_i'^T\hat{\bm{\gamma}}_c'(2)$
are invariant under change of boundary condition.
However, the more general procedure is somewhat $B_c$ dependent.

For the case of a constant Coulomb energy shift applied to all levels,
${\bm{\Delta}=\Delta\bm{1}}$, the commutator $[\bm{K},\bm{\Delta}]=0$.
Thus, in this particular situation, the procedure is exactly $B_c$ independent.
In the limit of a large number of levels, the various bases are complete
and the more general procedure would also be expected to become
$B_c$ independent.

\bibliography{spec-fac.bib}

\end{document}